\documentclass[letterpaper,twocolumn,10pt]{article}

\usepackage{zhanggroup}
\usepackage{caption}
\usepackage{color}
\usepackage{subcaption}
\usepackage{float}
\usepackage{tikz}
\usepackage{amsfonts}
\usepackage{xspace} 
\usepackage{amsmath}
\usepackage{mathrsfs}
\usepackage{amsmath}
\usepackage{amsthm}
\usepackage{subcaption}
\usepackage{booktabs}
\usepackage{multirow}
\usepackage{xcolor}
\usepackage{url}
\usepackage[absolute]{textpos}
\newcommand{\mypara}[1]{\smallskip\noindent{\bf {#1}} \xspace}
\newcommand{\RNum}[1]{\uppercase\expandafter{\romannumeral #1\relax}}
\definecolor{revision}{RGB}{0,0,255}

\newcommand{\revstart}{\begin{color}{revision}}
\newcommand{\revend}{~\!\!\end{color}}
\newcommand{\tabincell}[2]{\begin{tabular}{@{}#1@{}}#2\end{tabular}}

\begin{document}

\begin{textblock}{16}(1.9,1)
To Appear in 2021 ACM SIGSAC Conference on Computer and Communications Security, November 2021
\end{textblock}

\title{Membership Inference Attacks Against Recommender Systems}
\date{}

\author{
Minxing Zhang\textsuperscript{1,2}\thanks{These authors contributed equally to this work.}\ \ \ \
Zhaochun Ren\textsuperscript{1*}\thanks{Corresponding author.}\ \ \ \
Zihan Wang\textsuperscript{1*}\ \ \ \
Pengjie Ren\textsuperscript{1}\ \ \ \
\\
Zhumin Chen\textsuperscript{1}\ \ \ \
Pengfei Hu\textsuperscript{1}\ \ \ \
Yang Zhang\textsuperscript{2\dag}
\\
\textsuperscript{1}\textit{Shandong University}\ \ \ \textsuperscript{2}\textit{CISPA Helmholtz Center for Information Security}
}

\maketitle

\begin{abstract}
Recently, recommender systems have achieved promising performances and become one of the most widely used web applications.
However, recommender systems are often trained on highly sensitive user data, thus potential data leakage from recommender systems may lead to severe privacy problems.

In this paper, we make the first attempt on quantifying the privacy leakage of recommender systems through the lens of membership inference.
In contrast with traditional membership inference against machine learning classifiers, our attack faces two main differences.
First, our attack is on the user-level but not on the data sample-level.
Second, the adversary can only observe the ordered recommended items from a recommender system instead of prediction results in the form of posterior probabilities.
To address the above challenges, we propose a novel method by representing users from relevant items.
Moreover, a shadow recommender is established to derive the labeled training data for training the attack model.
Extensive experimental results show that our attack framework achieves a strong performance.
In addition, we design a defense mechanism to effectively mitigate the membership inference threat of recommender systems.\footnote{Our code is available at \url{https://github.com/minxingzhang/MIARS}.}
\end{abstract}

\section{Introduction}
\label{section:introduction}

As one of the most prevalent services in current web applications, recommender systems have been applied in various scenarios, such as online shopping, video sharing, location recommendation, etc.
A recommender system is essentially an information filtering system, relying on machine learning algorithms to predict user preferences for items.
One mainstream method in this space is collaborative filtering, which is based on traditional methods such as matrix factorization and latent factor model, predicting a user's preference from their historical behaviors combined with other users' similar decisions~\cite{HKR00, SFHS07}.
Another is the content-based recommendation~\cite{PB07,WF20}.
This approach aims to distinguish users' likes from dislikes based on their metadata (such as descriptions of the items and profiles of the users' preferences).
Recent advancement of deep learning techniques further boosts the performance of recommender systems~\cite{HLZNHC17}.

The success of recommender systems lies in the large-scale user data.
However, the data in many cases contains sensitive information of individuals, such as shopping preference, social relationship~\cite{BHPZ17}, and location information~\cite{STBH11}.
Recently, various research has shown that machine learning models, represented by machine learning classifiers, are prone to privacy attacks~\cite{SSSS17,PSMRTE18,NSH18,NSH19,SZHBFB19,MSCS19,SDSOJ19,SS19,JSBZG19,CYZF20,CTWJHLRBSEOR20,CTCP20,LZ21,HJBGZ21,NSTPC21}.
However, the privacy risks stemming from recommender systems have been left largely unexplored.

\begin{figure}[!t]
\centering
\includegraphics[width=0.95\columnwidth]{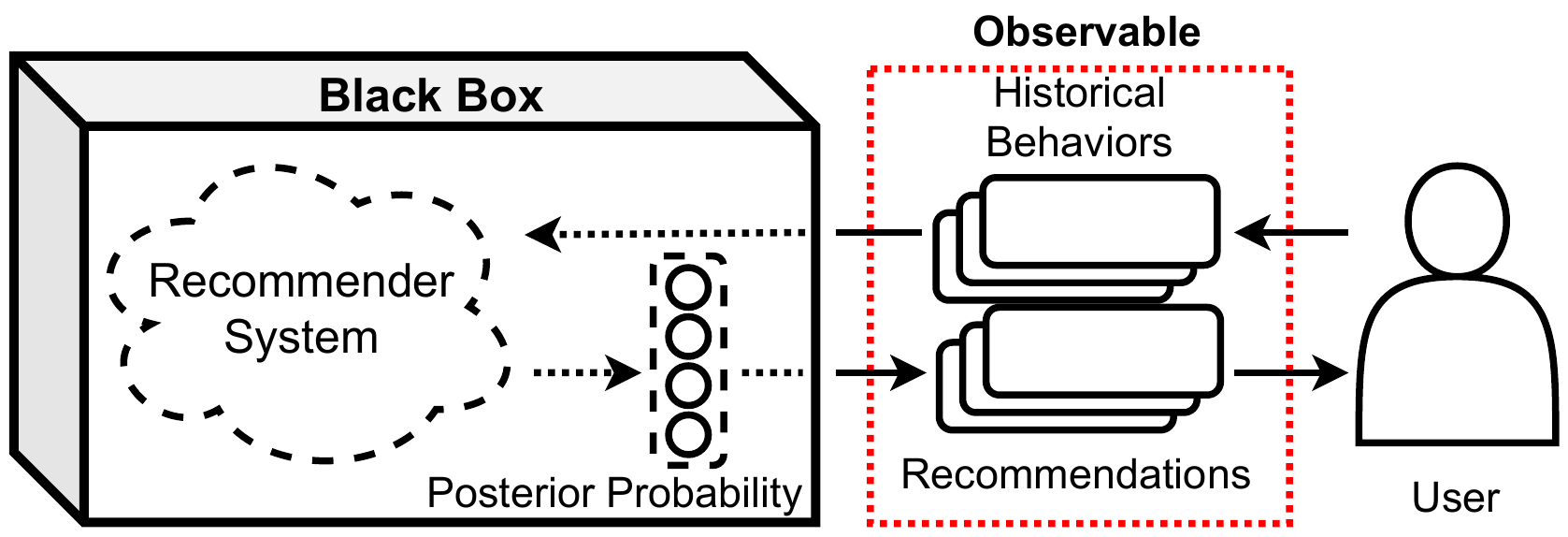}
\caption{An example of recommender systems.}
\label{figure:blackbox}
\end{figure} 

\subsection{Our Contributions}
\label{section:contribution}

In this paper, we take the first step quantifying the privacy risks of recommender systems through the lens of membership inference.
Compared to previous membership inference attacks against machine learning classifiers~\cite{SSSS17,SZHBFB19}, our attack faces two main differences.
First, the goal of our attack is to determine whether a user's data is used by the target recommender.
This indicates our attack is on the user-level while most of the previous attacks focus on the sample-level~\cite{SSSS17,SZHBFB19}.
Unlike sample-level membership inference, user-level membership inference has a broader scope as mentioned in previous works~\cite{SS19}, and it can help us gain a comprehensive understanding of recommender systems' privacy risks.
Second, from the adversary's perspective, only ranking lists of items are available from the target recommender which raises several technical challenges:
\begin{itemize}
    \item In our attack, as~\autoref{figure:blackbox} depicts, the adversary can only observe lists of items, even though recommender systems have already calculated posterior probabilities before making decisions.
    This setting is prevalent in the real world, such as the service provided by Amazon or Netflix.
    Besides, exposing less information can protect the intellectual property of recommendation service providers~\cite{SZHBFB19}.
    On the contrary, in classical membership inference attacks against classifiers, posterior probabilities used for decisions can be accessed by the adversary~\cite{SSSS17, SZHBFB19, SDSOJ19, NSH19}.
    \item Several recent membership inference attacks against classifiers focus on the decision-only (i.e., label-only) scenario~\cite{LZ21,CTCP20}.
    However, these studies either rely on the target model to label a shadow dataset~\cite{LZ21} or use adversarial examples~\cite{LZ21,CTCP20}, which are not practical when targeting recommender systems in the real world.
    Therefore, we aim for a new method to extract information from decision results for the attack model.
    \item Unlike classical classifiers, the outputs of recommender systems are ranking lists of items other than unordered labels.
    In that case, the order information plays an important role, and can substantially facilitate user preference predictions.
    Therefore, it is necessary for our attack model to capture order information from recommended items, which is still ignored by previous membership inference attack methods.
\end{itemize}

\mypara{Threat Model.}
The goal of the adversary is to conduct a membership inference attack against the target recommender system, inferring whether a target user's data is used to train the target recommender or not. However,
such attacks can lead to severe security and privacy risks.
Specifically, when a recommender system is trained with data from people with certain sensitive information, such as health status, knowing a user's data being part of the recommender's training data directly leaks their private information.
Moreover, membership inference also gains the adversary information about the target recommender's training data, which may jeopardize the recommender's intellectual property, since collecting high-quality data often requires a large amount of efforts~\cite{LHZG19, HJBGZ21}.
From a different angle, a user can also use membership inference as a tool to audit whether their data is used by the target recommender.

We assume the adversary has black-box access to the target recommender, the most difficult setting for the adversary~\cite{SSSS17}.
Instead of posterior probabilities for recommendations, only relevant items for users are available, such as rating or purchase and recommended items.
Due to the knowledge, a shadow recommender is established to derive labeled training data, for the attack model better inferring membership status in the target recommender.

\mypara{Attack Method.}
For user-level membership inference, we need to summarize each user's \textit{feature vector}, based on interactions between the target recommender and them, as the input to the attack model.
However, compared to the previous work of membership inference against classifiers, the adversary can only observe the recommended items from a recommender system instead of posterior probabilities as prediction results.
Thus, in the first step, the adversary constructs a user-item matrix for ratings with a dataset used to generate feature vectors.
Then, they factorize this matrix into two low-dimensional matrices, namely user matrix and item matrix.
Each item's feature vector can be represented by the corresponding row vector in the item matrix.
For each user, the adversary extracts two sets of items (one set contains items the user is recommended and the other contains the items the user interacted with) and calculates these two sets' center vectors, respectively.
The difference between these two center vectors for each user describes how accurate the recommender is for this user.
In that case, lower difference indicates a user's data is more likely to be used to train the recommender.
Therefore, we use this difference as the input to the attack model, i.e., user feature vector.
The adversary generates all the labeled training dataset for their attack model with the help of the shadow recommender.
To launch the attack, the adversary generates the target user's feature vector following the same steps and obtains the prediction from the attack model.

\mypara{Evaluation.}
To evaluate our attack, the adversary is assumed to have a shadow dataset that comes from the same distribution as the target recommender's training data, and know the target recommender's algorithm.
These assumptions are gradually relaxed based on our empirical evaluation.

Our experiments are performed on three benchmark recommendation datasets, i.e., the Amazon Digital Music (ADM)~\cite{HM16}, Lastfm-2k (lf-2k)~\cite{CBK11}, and Movielens-1m (ml-1m)~\cite{HK15}.
The recommendation algorithms we focus on include Item-Based Collaborative Filtering (Item), Latent Factor Model (LFM), and Neural Collaborative Filtering (NCF).
Evaluation results demonstrate that our attack is able to achieve an excellent performance.
\begin{itemize}
	\item In general, when the adversary knows the distribution of the target dataset and the target recommender's algorithm, the attack performance is extremely strong.
	For instance, when the target recommender uses NCF on the ADM dataset, our attack achieves an AUC of 0.987.
	Also, when the target algorithm is Item or NCF, our attack achieves better performances.
	\item When the adversary is not aware of the target recommender's algorithm, attack performances are reduced but still strong.
	For instance, on the lf-2k dataset, when the target recommender uses Item and the shadow recommender uses NCF, the attack performance decreases from an AUC of 0.929 to an AUC of 0.827.
	On the other hand, in some cases, the attack performances even increase.
	For instance, on the ml-1m dataset, when the target recommender uses Item and the shadow recommender uses LFM, the attack's AUC increases from 0.871 to 0.931.
	\item We further relax the assumption of the shadow dataset.
	Evaluation shows that even under such a scenario, our attack still achieves good performances in general.
	Note that, in some cases, when the adversary knows less about the target recommender, the attack even performs better.
	This demonstrates the good generalization ability of our attack.
\end{itemize}
In conclusion, the experimental results show the effectiveness of our attack, indicating that recommender systems are indeed vulnerable to privacy attacks.

\mypara{Defense.}
To mitigate the recommender's privacy risk, we propose a defense mechanism, namely \emph{Popularity Randomization}.
Popularity Randomization is deployed when the target recommender recommends items to its non-member users.
The normal strategy in such case is to provide non-member users with the most popular items.
However, to defend the attack, we enlarge the set of popular items, and randomly select a subset of them for recommendation.
Intuitively, while preserving the recommendation performance for non-member users, this approach enriches the randomness of the recommendation.

Experimental results show that Popularity Randomization can effectively mitigate membership inference.
For instance, Popularity Randomization (with a 0.1 ratio of recommendations to candidates) decreases the attack performances by more than $12\%$, $33\%$, and $41\%$ when the target algorithm is Item, LFM or NCF respectively.
Through analyses, we observe that Popularity Randomization has the greatest impact on the attack targeting on NCF.

\section{Method}
\label{section:method}

In this section, we first present some necessary definitions in~\autoref{section:definitions}, and then introduce the threat model for the membership inference attack against recommender systems in~\autoref{section:threatmodel}.
Next, we give overviews for recommender systems in~\autoref{section:recommendersystem} and our attack model in~\autoref{section:overview}.
Finally, we detail the proposed membership inference attack methods in~\autoref{section:MIA}.

\subsection{Definitions}
\label{section:definitions}
We present the following definitions for the attack process:
\begin{itemize}
    \item \textbf{Target Recommender}, trained on the \textbf{Target Dataset}, is the recommender system attacked by the adversary.
    \item \textbf{Shadow Recommender}, trained on the \textbf{Shadow Dataset}, is a recommender system built to infer the membership status of the target recommender and generate training data for the attack model.
    \item \textbf{Members} are the users whose data is used to train the recommender, while \textbf{Non-Members} are the ones whose data is not used.
    \item \textbf{Personalized Recommendation Algorithms} learn members' preferences from historical behaviors (such as purchases or ratings), which are also called \textbf{Interactions}.
    \textbf{Non-Personalized Recommendation Algorithms} are based on the predetermined rule, such as selecting the most popular or highest-rated items.
    According to different recommendation methods, members and non-members are provided with recommended items, which are also called \textbf{Recommendations}.
    \item \textbf{Feature Vectors} show the latent features, indicating item attributes or user preferences.
    \item \textbf{Attack Model} is used to infer whether the target user is a member, and  trained on the dataset generated from the shadow recommender.
\end{itemize}

\subsection{Threat Model}
\label{section:threatmodel}

\mypara{Adversary's Goal.}
The adversary aims to infer whether a user's data is used by a target recommender.
In fact, knowing a certain user's data being used by a recommender system directly leaks their private information.
Besides, knowing a user being part of the dataset can also allow the adversary to gain extra information of the target recommender's dataset.
This directly violates the intellectual property of the target recommender, since it is very expensive to collect high-quality training data.
Alternatively, membership inference can also be used as an auditing tool by a user to find out whether their data is used by the target recommender.

\mypara{Adversary's Knowledge.} 
We assume an adversary has only black-box access to the target recommender. That is, adversary can only observe the items recommended to a target user (i.e., \emph{recommendations}), and the user's history (i.e., \emph{interactions}), such as rating and purchase, instead of posterior probabilities for recommendation predictions.
In that case, the adversary needs to profile users by their interactions and recommendations.
Meanwhile, a shadow recommender is built to generate labeled data for the attack model, since ground truth membership is unavailable from the target recommender.

\subsection{Recommender Systems}
\label{section:recommendersystem}

In this section, the framework of recommender systems is briefly introduced.

Recommendation algorithms output recommended items based on the information learnt from input.
In the paper, two types of recommendation algorithms are mainly involved: personalized and non-personalized recommendation algorithms.
For members, items are recommended according to the preferences of members.
Meanwhile, lacking non-members' data, non-personalized recommendation algorithms are conducted, and provide most popular items for non-members.
Specifically, Item-Based Collaborative Filtering (Item)~\cite{SKKR01}, Latent Factor Model (LFM) and Neural Collaborative Filtering (NCF)~\cite{HLZNHC17} are adopted as the personalized recommendation algorithms for members.
As for the non-personalized recommendation algorithm, the most popular items are provided to non-members, which is also called the popularity recommendation algorithm in the paper. We briefly introduce the above algorithms as follows:
\begin{itemize}
    \item \textbf{Item} calculates the similarity between items aiming to find the ones which are closed to users' likes.
    \item \textbf{LFM} builds a latent space to bridge user preferences and item attributes.
    \item \textbf{NCF} combine the deep learning technology with collaborative filtering to enhance the recommendation performances.
    \item Users are provided with the most popular items by the \textbf{popularity recommendation algorithm}.
\end{itemize}

In general, a recommender system $\mathcal{A}_{RS}$ learns user preferences from the interactions, sometimes with the external knowledge (such as gender and location information) for users.
According to the predicted preferences, the recommender system provides users with multiple items.
This procedure can be formulated as:
$$\mathcal{A}_{RS}: (\mathcal{I}_{RS}, \mathcal{K}_{RS}) \rightarrow \mathcal{R}_{RS},$$
where $\mathcal{A}_{RS}$ is a recommender system learning the preferences of users from their interactions $\mathcal{I}_{RS}$ and the external knowledge $\mathcal{K}_{RS}$.
And $\mathcal{R}_{RS}$ denotes recommended items to users.
In the paper, we mainly use the interactions of users.
Thus, we define a recommender system as:
$$\mathcal{A}_{RS}: \mathcal{I}_{RS} \rightarrow \mathcal{R}_{RS},$$
where $\mathcal{I}_{RS}$ is a set of lists of interactions for users and $\mathcal{R}_{RS}$ is a set of ordered lists of recommendations for users.
Concretely, $\mathcal{I}_{RS} = \{L_{I}^{n}\}_{n=1}^{N_{u}}$ and $\mathcal{R}_{RS} = \{L_{R}^{n}\}_{n=1}^{N_{u}}$, where $L_{I}^{n}$ is the list of interactions and $L_{R}^{n}$ is the ordered list of recommendations for the $n^{th}$ user, and $N_{u}$ is the number of users.

\subsection{Attack Overview}
\label{section:overview}

\begin{figure}[!t]
\centering
\includegraphics[width=\columnwidth]{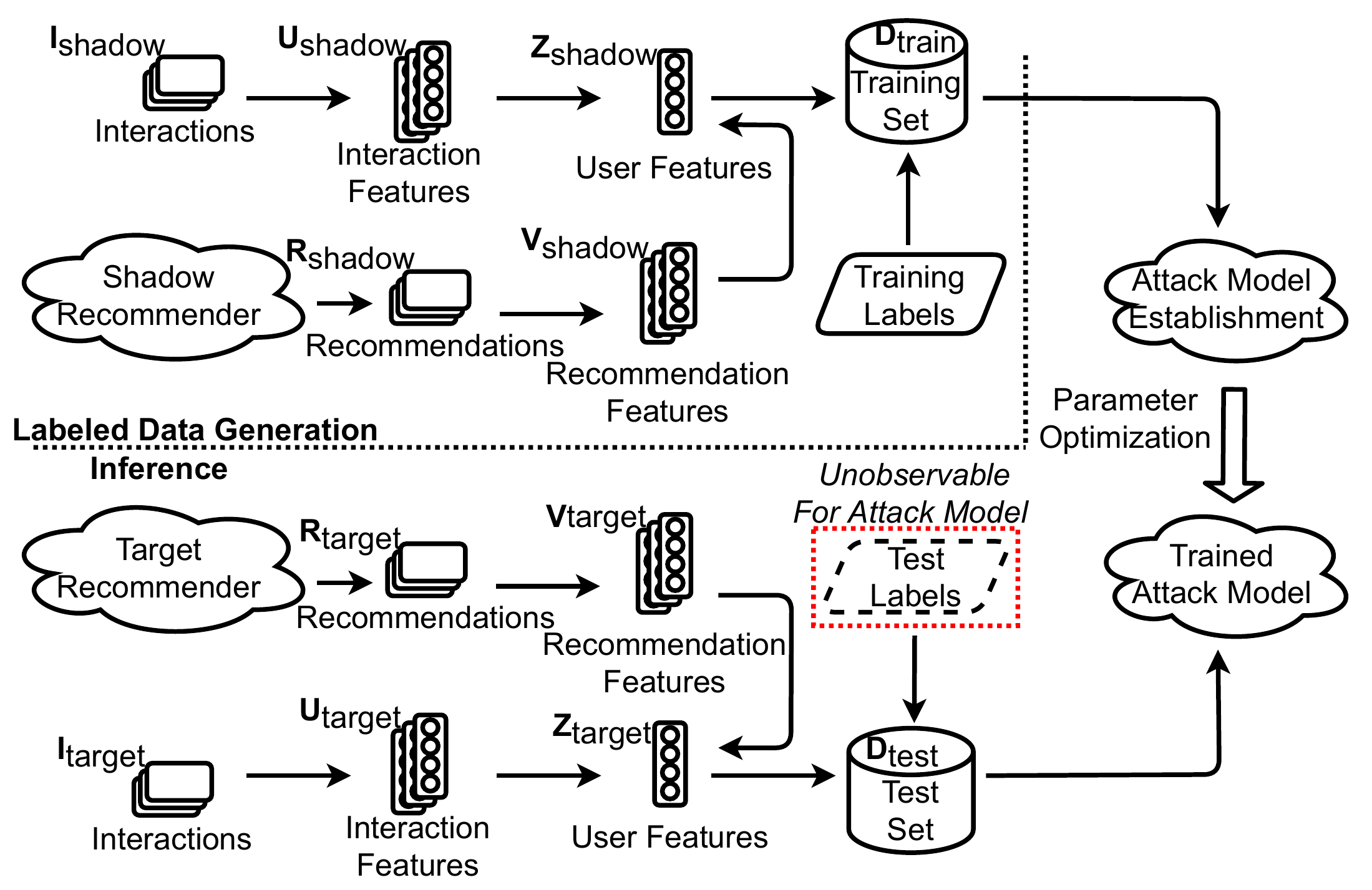}
\caption{The framework of the membership inference attack against a recommender system.}
\label{figure:attackframework}
\end{figure} 

In this section, we give an overview of our attack.
As~\autoref{figure:attackframework} demonstrated, the attack process follows three steps: Labeled Data Generation, Attack Model Establishment, and Parameter Optimization.

\mypara{Labeled Data Generation.}
To represent items, an item matrix is derived, by factorizing a user-item rating matrix.
Due to the black-box access to the target recommender for the adversary, a shadow recommender is built to generate labeled training data for the attack model.
Moreover, we represent interactions and recommendations of users using corresponding feature vectors.
After that, a user is profiled by the difference between two centers of their interactions and recommendations.
And each user is labeled with 1 or 0, indicating they are a member or non-member.

\mypara{Attack Model Establishment.}
Inspired by~\cite{SZHBFB19}, a two-hidden-layer Multi-Layer Perceptron (MLP) is utilized as the attack model $A_{attack}$ to infer membership status.
Each hidden layer is followed by a \rm{ReLU} activation layer.
And a \rm{softmax} function is used as the output layer to predict the probability of the membership.

\mypara{Parameter Optimization.}
After \emph{Labeled Data Generation} and \emph{Attack Model Establishment}, as shown in~\autoref{figure:attackframework}, the adversary updates parameters of the attack model.
In the inference stage, the test dataset for target users are established following the same steps as training data generation.
The membership status for target users is inferred by the trained attack model.

\subsection{Membership Inference Attack}
\label{section:MIA}

In this section, we detail our proposed membership inference attack against a recommender system.
As mentioned before, the whole attack consists of three steps to achieve the adversary's goal: Labeled Data Generation, Attack Model Establishment, and Parameter Optimization.

\mypara{Labeled Data Generation.}
Training data is required during the $\mathcal{A}_{attack}$ optimization process.
However, the adversary cannot obtain membership status directly from the target recommender $\mathcal{A}_{target}$.
To address this problem, a shadow recommender $\mathcal{A}_{shadow}$ is developed to mimic the dataset and recommendation algorithm of the target recommender.

As mentioned in~\autoref{section:contribution}, only recommended item lists from target recommender systems can be observed.
Inspired by the previous works~\cite{K08, HZKC16}, \emph{matrix factorization} is adopted to project users and items into a shared latent space.
Specifically, a $p \times q$ user-item matrix $\mathbf{M}^{f}$ is built using ratings of users to items, where $p$ and $q$ are the number of users and items respectively.
Values in $\mathbf{M}^{f}$ are ratings ranging from 1 to 5, indicating how much users prefer these items.
Then, $\mathbf{M}^{f}$ is factorized into two low-dimensional matrices, namely latent user matrix $\mathbf{M}^{user} \in \mathcal{R}^{p \times l}$ and latent item matrix $\mathbf{M}^{item} \in \mathcal{R}^{q \times l}$, where we denote $l$ as the dimension of the latent feature space.
We apply matrix factorization to find optimized $\mathbf{M}^{user}$ and $\mathbf{M}^{item}$ by minimizing the loss function $L_{MF}$:
$$L_{MF} = \left\|\mathbf{M}^{f}-\hat{\mathbf{M}}^{f}\right\|_2 \quad where\quad\hat{\mathbf{M}}^{f} = \mathbf{M}^{item} \cdot {\mathbf{M}^{user}}^\mathrm{T},$$
where $\hat{\mathbf{M}}^{f}$ is a predicted user-item matrix which contains the predicted scores of users rating items.
Besides, $\mathbf{M}^{user}$ and $\mathbf{M}^{item}$ present the predicted preferences of users and the predicted attributes of items, respectively.
Each row of the item matrix $\mathbf{M}^{item}$ represents the feature vector of the corresponding item.
Note that, since $\mathbf{M}^{user}$ may not cover all users in the shadow and target recommenders, $\mathbf{M}^{user}$ is not used to represent users.

To this end, training data for the attack model can be generated from $\mathcal{A}_{shadow}$.
The shadow dataset $\mathcal{D}_{shadow}$ are split into two disjoint sets for members and non-members, which are denoted by $\mathcal{D}_{shadow}^{in}$ and $\mathcal{D}_{shadow}^{out}$, respectively.
These datasets are composed of 3-tuples in the form of 
$(uID, iID, score)$, indicating scores rated by users to items.
For instance, a 3-tuple $(2, 3, 4)$ in datasets means that the $2^{nd}$ user rates the $3^{rd}$ item a score of $4$.
Ratings in $\mathcal{D}_{shadow}^{in}$ and $\mathcal{D}_{shadow}^{out}$ can be seen as interactions of users to items, and sets of interaction lists for members and non-members can be obtained, denoted as $\mathcal{I}_{shadow}^{in}$ and $\mathcal{I}_{shadow}^{out}$, respectively. In that case, 
each user has a list of interactions. For example, if a user rates the $2^{nd}$, $4^{th}$, $6^{th}$ and $8^{th}$ items, the corresponding interaction list is $\{2, 4, 6, 8\}$.

Next, $\mathcal{A}_{shadow}$ is established to mimic $\mathcal{A}_{target}$, and provides users with recommendations according to their preferences. The sets of recommendation lists for members and non-members are denoted by $\mathcal{R}_{shadow}^{in}$ and $\mathcal{R}_{shadow}^{out}$, respectively.
Similar as interactions, each user is associated with a list of recommendations. However, $\mathcal{R}_{shadow}^{in}$ and $\mathcal{R}_{shadow}^{out}$ are sets of \emph{ordered} lists of recommendations.
Formally, the recommendation process can be formulated as follows:
$$\mathcal{A}_{shadow}: f_{per}(\mathcal{I}_{shadow}^{in}) = \mathcal{R}_{shadow}^{in}$$
$$\mathcal{A}_{shadow}: f_{pop}(\mathcal{I}_{shadow}^{in}) = \mathcal{R}_{shadow}^{out},$$
where $f_{per}$ performs a personalized recommendation algorithm based on the behaviors of members.
Meanwhile, since non-members' data is unavailable to $\mathcal{A}_{shadow}$, $f_{pop}$ performs the popularity recommendation algorithm (a non-personalized recommendation algorithm) based on the statistical results from $\mathcal{I}_{shadow}^{in}$.
Besides, $\mathcal{I}_{shadow}^{in}$ is a set of lists of interactions for members, and $\mathcal{R}_{shadow}^{in}$ and $\mathcal{R}_{shadow}^{out}$ are sets of ordered lists of recommended items for members and non-members respectively.

Using item feature representations, we can vectorize the interaction and recommendation sets as follows:
$$\mathcal{I}_{shadow}\stackrel{vec}{\longrightarrow}\mathcal{U}_{shadow}$$
$$\mathcal{R}_{shadow}\stackrel{vec}{\longrightarrow}\mathcal{V}_{shadow},$$
where $\mathcal{U}_{shadow}$ and $\mathcal{V}_{shadow}$ are sets of lists of the feature vectors for the corresponding items in $\mathcal{I}_{shadow}$ and $\mathcal{R}_{shadow}$.

Given that each user has a list of interactions and is provided with an ordered list of recommendations, the adversary is able to represent users by their relevant items.
To be specific, for the $i^{th}$ user, the representation is generated with the following two steps:
\begin{itemize}
    \item [1)] 
    Center vectors of the interactions' and recommendations' feature vectors of the $i^{th}$ user are calculated:
    $$\overline{\mathcal{U}}_i = \sum\limits_j \mathcal{U}_{i,j} / N_i^{int}$$ $$\overline{\mathcal{V}}_i = \sum\limits_j \mathcal{V}_{i,j} / N_i^{rec},$$
    where $\overline{\mathcal{U}}_i$ and $\overline{\mathcal{V}}_i$ are the center vectors of the feature vectors for the interactions and recommendations of the $i^{th}$ user, and $N_i^{int}$ and $N_i^{rec}$ are the corresponding quantities.
    Besides, $\mathcal{U}_{i,j}$ and $\mathcal{V}_{i,j}$ are the feature vectors for the $j^{th}$ interaction and recommendation of the $i^{th}$ user, respectively.
    \item [2)]
    The difference between the two center vectors are obtained: 
    $$\mathbf{z}_i = \overline{\mathcal{U}}_i - \overline{\mathcal{V}}_i.$$
    In the paper, we employ $\mathbf{z}_i$ as the feature vector for the $i^{th}$ user, which takes not only the user's history but also the predicted preference into consideration.
\end{itemize}
Meanwhile, each user is assigned a label of 1 or 0, indicating their membership (i.e., 1 means member and 0 means non-member).
The training dataset $\mathcal{D}_{train} = \{(feature_{i}, label_{i})\}_{i=1}^{N}$ contains feature vectors and labels of all users, where the pair $(feature_{i}, label_{i})$ denotes the feature vector and label for the $i^{th}$ user.

\mypara{Attack Model Establishment.}
Inspired by~\cite{SZHBFB19}, a MLP is established as the attack model $\mathcal{A}_{attack}$.
The output of $\mathcal{A}_{attack}$ is a 2-dimension vector representing probabilities for the membership status.
For the $i^{th}$ user, the prediction can be formulated as follows:
$$\mathbf{h}_{1} = \rm{ReLU}(\mathbf{W}_{1}\mathbf{z}_{i}+\mathbf{b}_{1})$$
$$\mathbf{h}_{2} = \rm{ReLU}(\mathbf{W}_{2}\mathbf{h}_{1}+\mathbf{b}_{2})$$
$$\mathbf{y}_{i} = \rm{softmax}(\mathbf{h}_{2}),$$
where $\mathbf{z}_{i}$ is the input of $\mathcal{A}_{attack}$ as well as the $i^{th}$ user's feature vector in our attack.
And $\mathbf{W}_{1}$, $\mathbf{W}_{2}$, $\mathbf{b}_{1}$ and $\mathbf{b}_{2}$ are the parameters updated in the training process.
$\rm{ReLU}(\cdot)$ is an activation function working on the outputs of two hidden layers, and $\rm{softmax}(\cdot)$ is used for normalization which is required by the cross-entropy loss.
Besides, $\mathbf{h}_{1}$ and $\mathbf{h}_{2}$ are the results of two hidden layers after $\rm{ReLU}(\cdot)$.
And $\mathbf{y}_{i}$ is the predicted result for the input $\mathbf{z}_{i}$, which is a 2-dimension vector indicating the possibilities of $\mathbf{z}_{i}$ belonging to members and non-members, respectively.

\mypara{Parameter Optimization.}
In this section, the parameter optimization process for the attack model is described.
Stochastic gradient descent is adopted to update parameters, aiming to minimize the cross-entropy loss function $L_{MLP}$:
$$L_{MLP} = - \sum\limits_{i=1}^{N_{train}} ( \mathbf{y}^{*}_{i}\rm{log}\mathbf{y}^{\prime}_{i} + (1-\mathbf{y}^{*}_{i})\rm{log}(1-\mathbf{y}^{\prime}_{i}) ),$$
where $\mathbf{y}^{*}_{i}$ is the ground truth label for the $i^{th}$ target user.
And $\mathbf{y}^{\prime}_{i}$ is the predicted possibility of the $i^{th}$ target user belonging to members.
Besides, $N_{train}$ is the size of training data.

Test data $\mathcal{D}_{test}$ for the attack model is generated from the target recommender in the same way as the training data. The trained attack model $\mathcal{A}_{attack}^{\prime}$ conduct a prediction given a target user feature vector $\mathbf{z}_{target}$, i.e., $\mathcal{A}_{attack}^{\prime}(\mathbf{z}_{target}) = \mathbf{y}_{target}$, where $\mathbf{y}_{target} = \left( \begin{array}{c} a\\ b\\ \end{array} \right)$ is a 2-dimension vector, and the values of $a$ and $b$ indicate the probabilities that the target user belongs to non-members and members respectively.
According to the predicted results, the adversary infers the membership status of the target user.
Concretely, when $a < b$, the target user is predicted to be a member.
Otherwise, they are predicted to be a non-member. 

\section{Experiments}
\label{section:experiment}

In this section, we first demonstrate experimental setup, including recommendation methods, datasets, preprocessing process, evaluation metrics, implementation details, and notations in~\autoref{section:experimentalsetups}.
Then, we evaluate the performances of original recommender systems in~\autoref{section:recommendationperformance}. 
Moreover, we investigate membership inference attacks against recommender systems in~\autoref{section:attackperformance} and conduct detailed analyses on the influences of hyperparameters in~\autoref{section:hyperparameters}.
Finally, we present extensive analysis to comprehensively investigate the attack model in~\autoref{section:extensiveanalysis}.

\subsection{Experimental Setup}
\label{section:experimentalsetups}

\mypara{Recommendation Methods.}
Personalized recommendation algorithms are adopted for members, including Item-Based Collaborative Filtering (Item)~\cite{SKKR01}, Latent Factor Model (LFM) and Neural Collaborative Filtering (NCF)~\cite{HLZNHC17}.
Meanwhile, due to the lack of non-members' data, a recommender system provides non-members with the most popular items, which is named the \emph{popularity recommendation algorithm} in our paper.

\mypara{Datasets.}
We utilize three real-world datasets in our experiments, including Amazon Digital Music (ADM)~\cite{HM16}, Lastfm-2k (lf-2k)~\cite{CBK11}, and Movielens-1m (ml-1m)~\cite{HK15}, to evaluate our attack strategies.
All these datasets are commonly-used benchmark datasets for evaluating recommender systems.
Note that only ratings in these datasets are used for our evaluation in the experiments.
Scores range from 1 to 5, which indicates how much users like musics (ADM and lf-2k) or movies (ml-1m).

\mypara{Preprocessing.}
For each dataset, we divide it into three disjoint subsets, i.e. a shadow dataset, a target dataset and a dataset for extracting item features.
Then, the following processing methods are implemented to these subsets:
\begin{itemize}
    \item To generate feature vectors for users, the dataset for item feature should contain all items of the target and shadow recommenders.
    \item For the shadow or target dataset, we further divide it into two disjoint parts, which are used to conduct recommendations to members and non-members, respectively.
    Moreover, following the previous work~\cite{HLZNHC17}, we filter out the users who have less than 20 interactions.
\end{itemize}
In our experiments, recommender systems conduct recommendations based on implicit feedback.
We assign values of 1 to the user-item pairs when there exist interactions between these users and items. And other user-item pairs are assigned 0.
In LFM and NCF, recommender systems require both positive and negative instances. We randomly sample negative user-item pairs from the pairs scoring 0 and regard the pairs assigned 1 as positive instances.
We keep the same number of negative samples as positive samples for the dataset balance.

\mypara{Evaluation Metrics.}
We use AUC (area under the ROC curve) as the metric to evaluate attack performances.
Following the definition of the attack, we regard members as \emph{positive data points} and non-members as \emph{negative data points}. AUC indicates the proportion of the prediction results being positive to negative.
For example, if the attack model utilizes Random Guess to conduct a membership inference, the AUC is close to 0.5.

\mypara{Implementation Details.}
We build a MLP with 2 hidden layers as the attack model.
The first hidden layer has 32 units and the second layer has 8 units, both followed by a ReLU layer.
And we utilize a softmax layer as the output layer.
For the optimizer, we employ Stochastic Gradient Descent (SGD) with a learning rate of 0.01 and a momentum of 0.7.
Besides, we use cross entropy as the loss function and the model is trained for 20 epochs.

In the paper, members are recommended by Item, LFM and NCF while non-members are recommended by the popularity recommendation algorithm.
Note that, Item and the popularity recommendation algorithm do not need the iterative process of updating parameters.
The detailed model configurations of LFM and NCF are shown as follows:
\begin{itemize}
    \item \textbf{LFM.} We adopt the SGD algorithm to update parameters with a learning rate of 0.01 and conduct LFM with a regularization coefficient of 0.01 to enhance the model's generalization ability.
    Then we train the model for 20 epochs.
    \item \textbf{NCF.} We use Adam as the optimizer with a learning rate of 0.001.
    And we build the MLP part with 3 hidden layers containing 64, 32 and 16 hidden units respectively.
    Meanwhile, the embedding size of the Generalized Matrix Factorization (GMF) part is 8~\cite{HLZNHC17}.
    In addition, the number of negative samples corresponding to per positive sample is set to 4.
    Then we train the model for 20 epochs with a batch size of 256.
\end{itemize}

\begin{table}[!t]
\centering
\caption{Notations for different settings. ``$*$'' stands for any algorithm or dataset used to construct or train the shadow or target model.}
\label{table:notation}
\setlength{\tabcolsep}{2mm}{
    \begin{tabular}{ l | l }
    \toprule
    \makebox Notation & Illustrations\\
    \midrule
    A$*$ & \tabincell{l}{Trained on the ADM dataset.}\\
    \hline
    L$*$ & \tabincell{l}{Trained on the lf-2k dataset.}\\
    \hline
    M$*$ & \tabincell{l}{Trained on the ml-1m dataset.}\\
    \hline
    $*$I & \tabincell{l}{ Implemented by Item algorithm.}\\
    \hline
    $*$L & \tabincell{l}{ implemented by LFM algorithm.}\\
    \hline
    $*$N & \tabincell{l}{ implemented by NCF algorithm.}\\
    \hline
    AI$**$ & \tabincell{l}{The shadow recommender is implemented\\by Item algorithm on the ADM dataset.}\\
    \hline
    $**$AI & \tabincell{l}{The target recommender is implemented\\by Item algorithm on the ADM dataset.}\\
    \hline
    AIMN & \tabincell{l}{The shadow recommender is implemented\\by Item algorithm on the ADM dataset, and\\the target recommender is implemented by\\NCF algorithm on the ml-1m dataset.}\\
    \bottomrule
    \end{tabular}
}
\end{table}

\mypara{Notations.}
To clarify the experimental settings, notations are demonstrated in~\autoref{table:notation}, where ``$*$'' stands for any algorithm or dataset used to construct or train the shadow or target model. For example, ``A$*$'' could be the combination of ``ADM$+$Item'', ``ADM$+$LFM'' or ``ADM$+$NCF''. 
Note that, not all possible combinations are listed due to the space limit.

In the experiments, there are two kinds of combinations in the paper (i.e., 2-letter and 4-letter combinations).
For the 2-letter combinations, the first letter, i.e., ``A,'' ``L'' or ``M'', indicates the shadow (or target) dataset, and the second letter, i.e., ``I,'' ``L'' and ``N'', indicates the recommendation algorithm.
For the 4-letter combinations, the first two letters represent the dataset and algorithm of the shadow recommender and the last two letters denote the dataset and algorithm of the target recommender.
For instance, ``AIMN'' means that the adversary establishes a shadow recommender with Item on the ADM dataset to attack a target recommender implemented by NCF on the ml-1m dataset.

\subsection{Recommendation Performance}
\label{section:recommendationperformance}

\begin{figure}[!t]
\centering
\includegraphics[width=0.95\columnwidth]{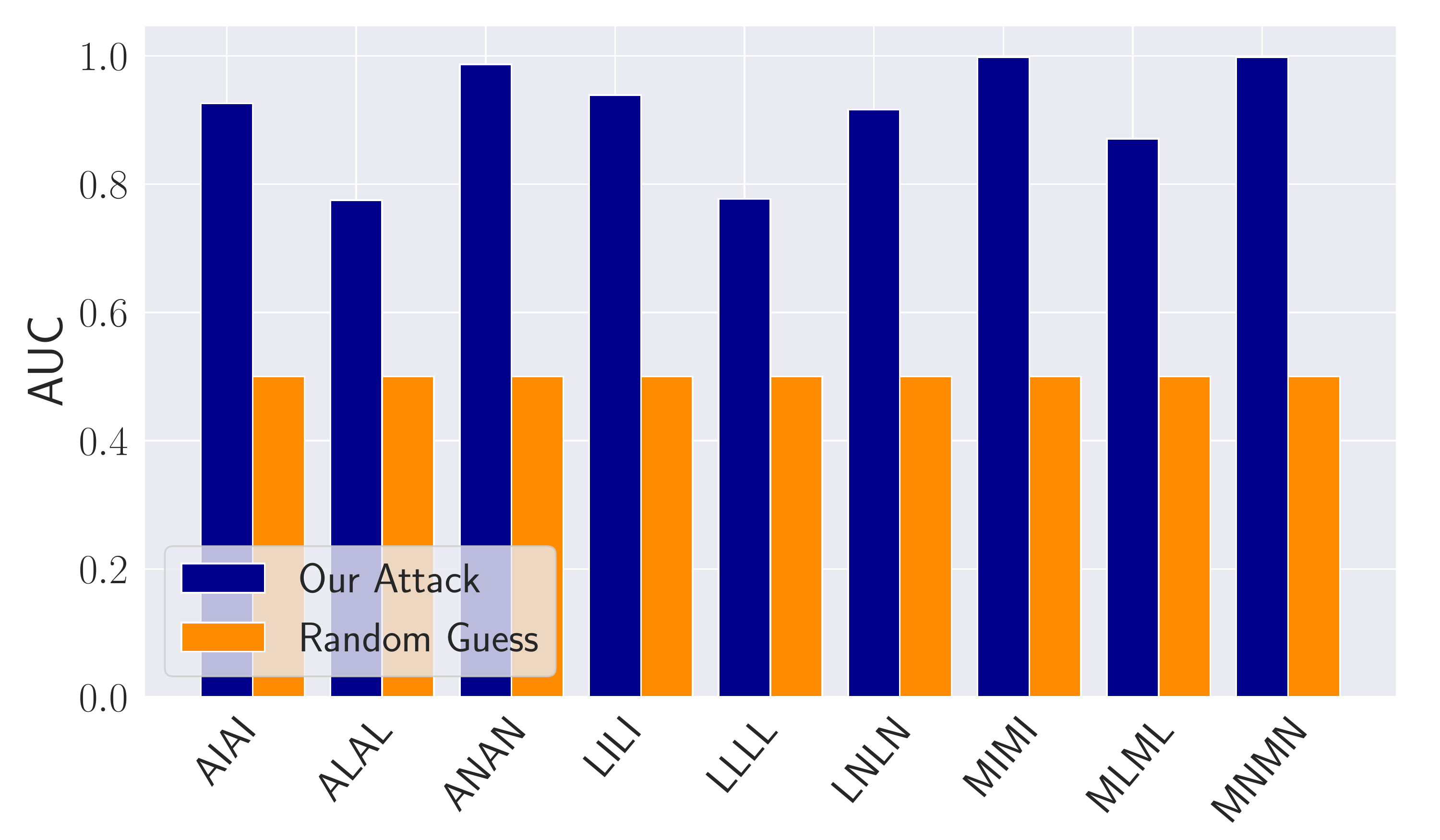}
\caption{The attack performances under the assumption \RNum{1}.}
\label{figure:SameAlgorithhmSameDataset}
\end{figure} 

\begin{table}[!t]
\centering
\caption{The HR@100 of the shadow and target recommenders.}
\label{table:rs_perform}
\setlength{\tabcolsep}{4.3mm}{
\begin{tabular}{l | c   c   c}
\toprule
\makebox ADM & Item & LFM & NCF\\
\midrule
shadow & 0.222 & 0.119 & 0.116\\
target & 0.224 & 0.204 & 0.123\\
\bottomrule
\end{tabular}
\begin{tabular}{l | c   c   c}
\makebox lf-2k & Item & LFM & NCF\\
\midrule
shadow & 0.652 & 0.478 & 0.625\\
target & 0.650 & 0.468 & 0.637\\
\bottomrule
\end{tabular}
\begin{tabular}{l | c   c   c}
\makebox ml-1m & Item & LFM & NCF\\
\midrule
shadow & 0.943 & 0.856 & 0.721\\
target & 0.951 & 0.860 & 0.713\\
\bottomrule
\end{tabular}
}
\end{table}
We adopt HR@$k$ as the metric to evaluate the recommendation performance, where $k=100$ is consistent with the experimental setting and Hit Rate (HR) presents the proportion of recommendations including the ground truth.
We can see from the results in~\autoref{table:rs_perform} that, in general, recommender systems achieve the best performance on the ml-1m dataset.
Specifically, the shadow recommender obtains a hit rate of 0.856 when using LFM on the ml-1m dataset.

\subsection{Attack Performance}
\label{section:attackperformance}

We perform experiments on the ADM, lf-2k and ml-1m datasets with three typical recommendation algorithms, including Item, LFM and NCF.
Experimental results show that our method is able to achieve strong attack performances.
We draw the following conclusions:

\begin{figure*}[!t]
\centering
\begin{subfigure}{\columnwidth}
\centering
\includegraphics[width=0.45\columnwidth]{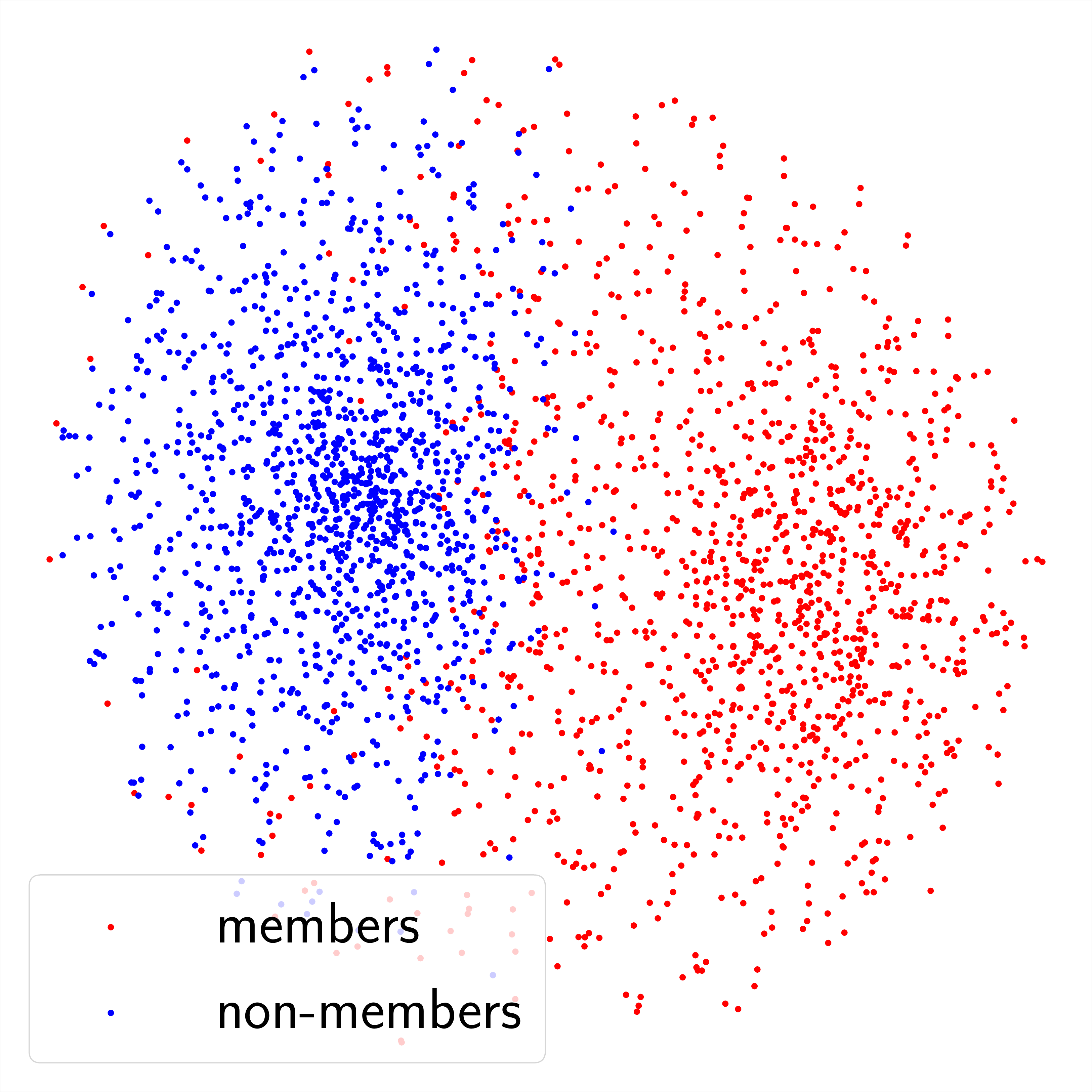}
\includegraphics[width=0.45\columnwidth]{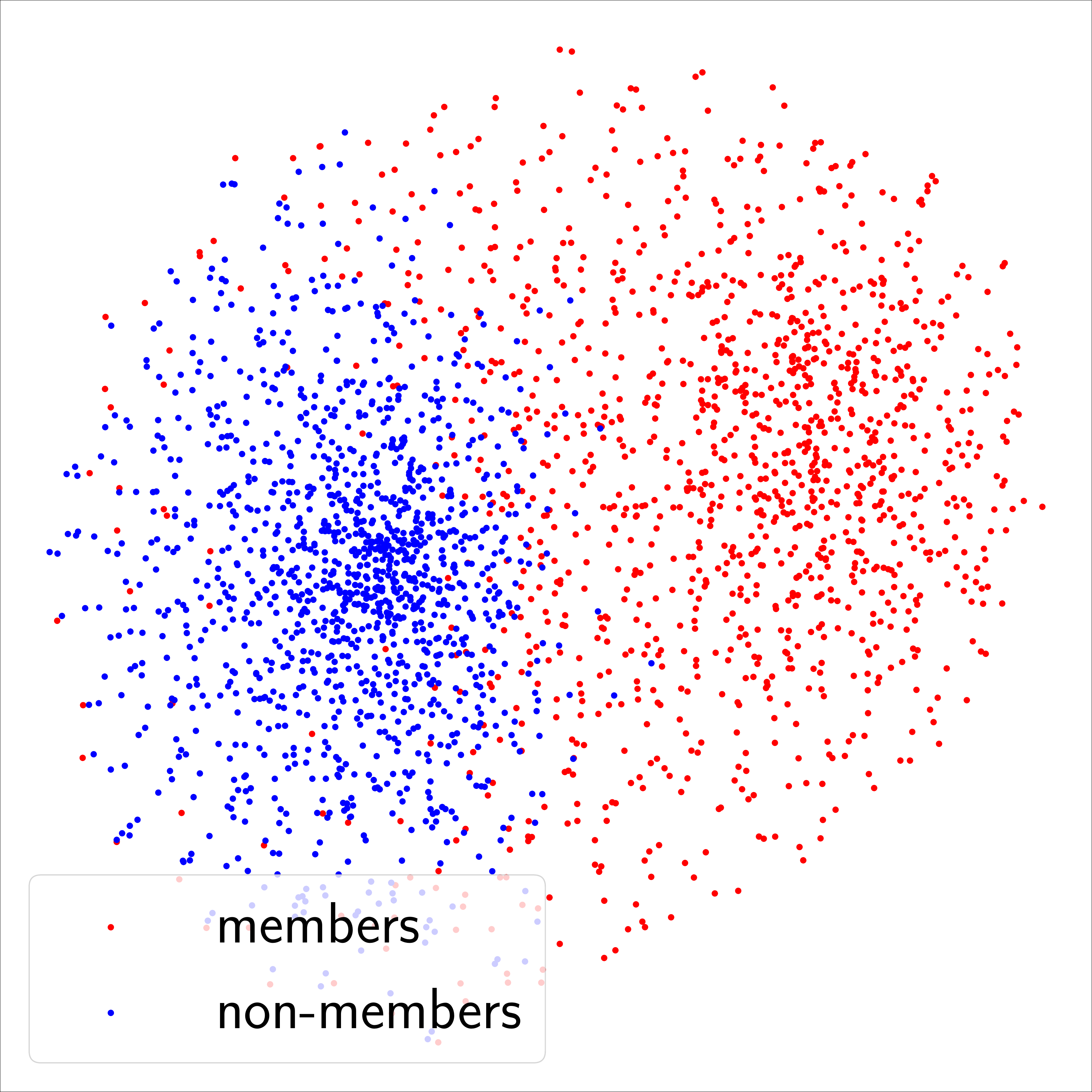}
\caption{ADM\_NCF\_shadow (left) and ADM\_NCF\_target (right)}
\label{figure:tsne_ANAN}
\end{subfigure}
\begin{subfigure}{\columnwidth}
\centering
\includegraphics[width=0.45\columnwidth]{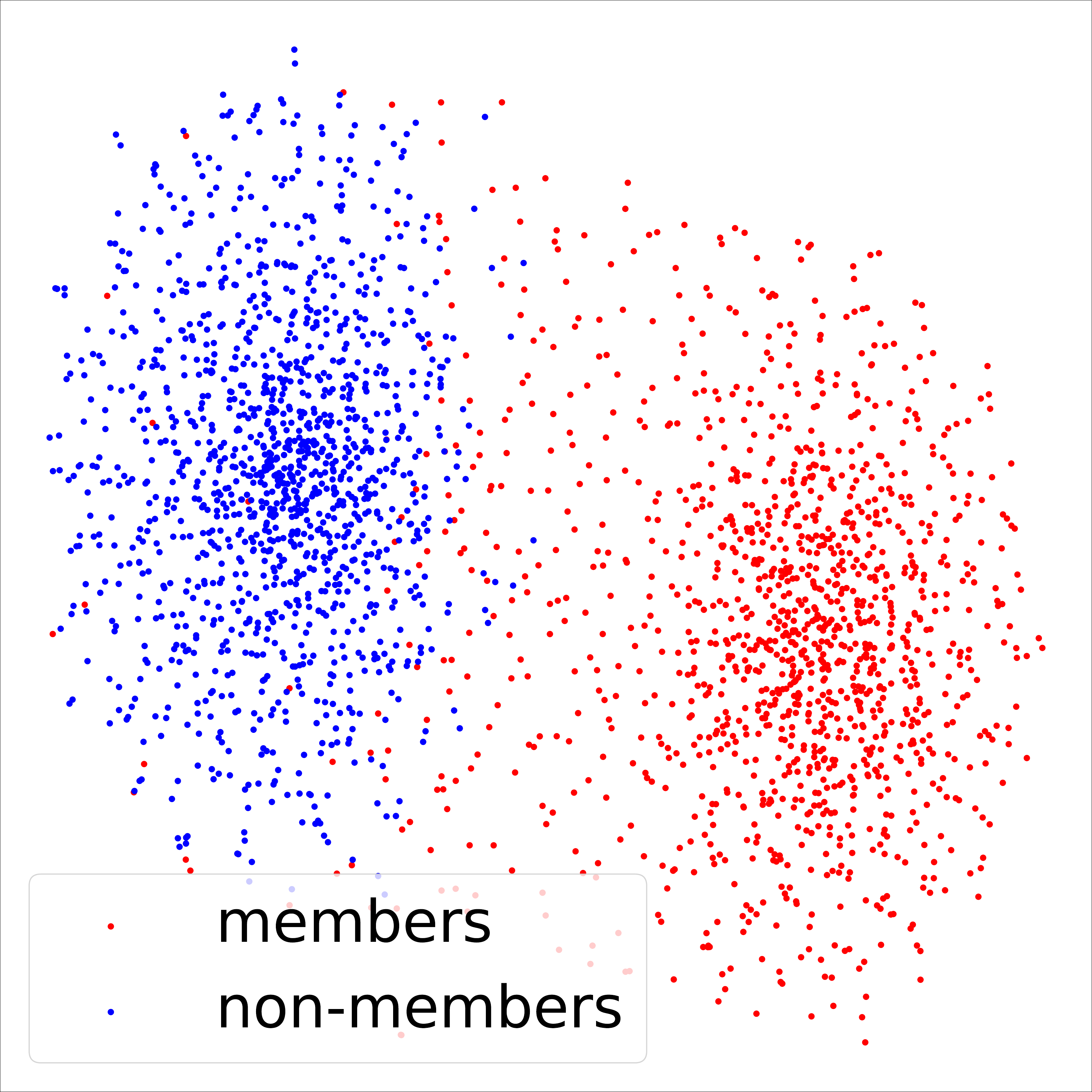}
\includegraphics[width=0.45\columnwidth]{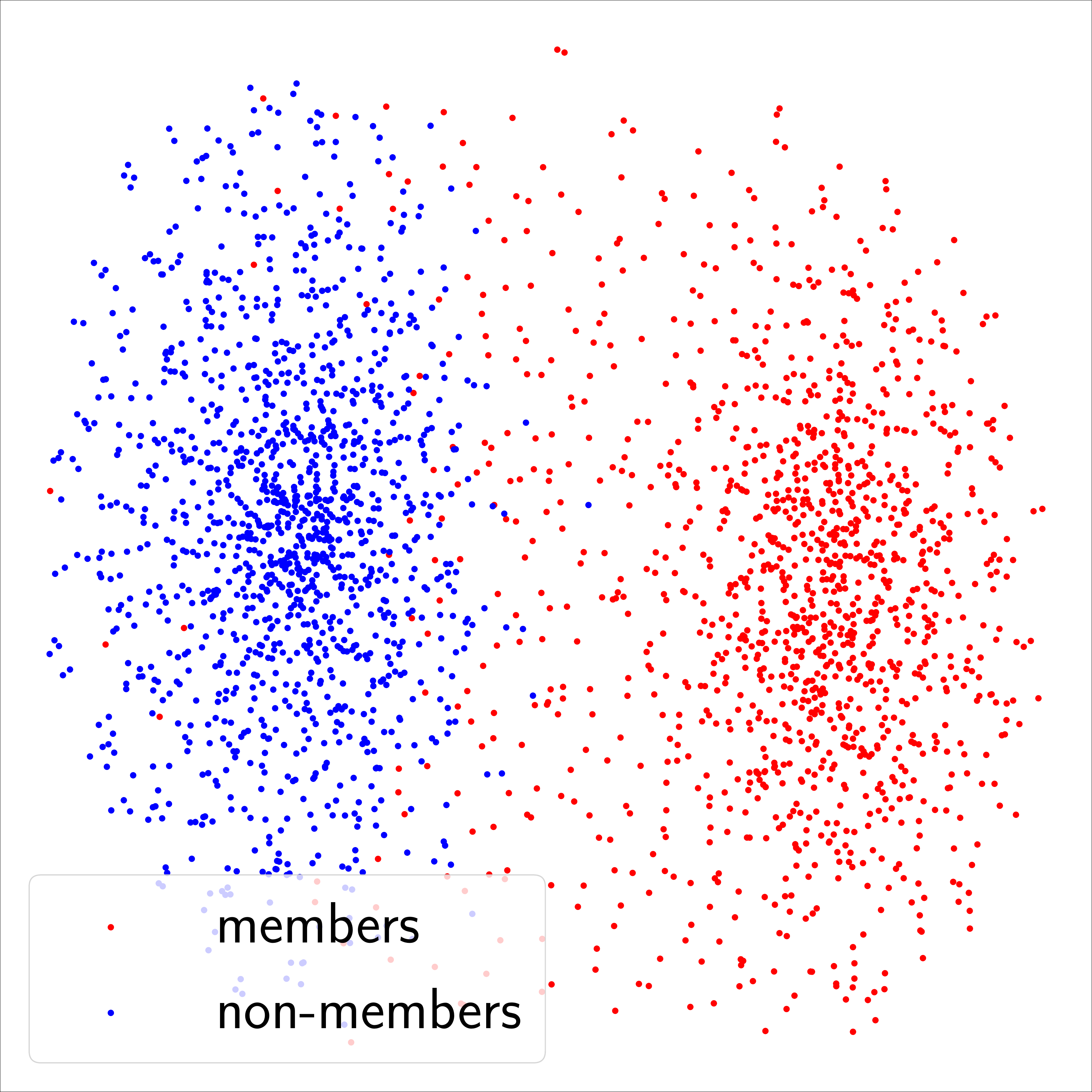}
\caption{ADM\_LFM\_shadow (left) and ADM\_LFM\_target (right)}
\label{figure:tsne_ALAL}
\end{subfigure}
\begin{subfigure}{\columnwidth}
\centering
\includegraphics[width=0.45\columnwidth]{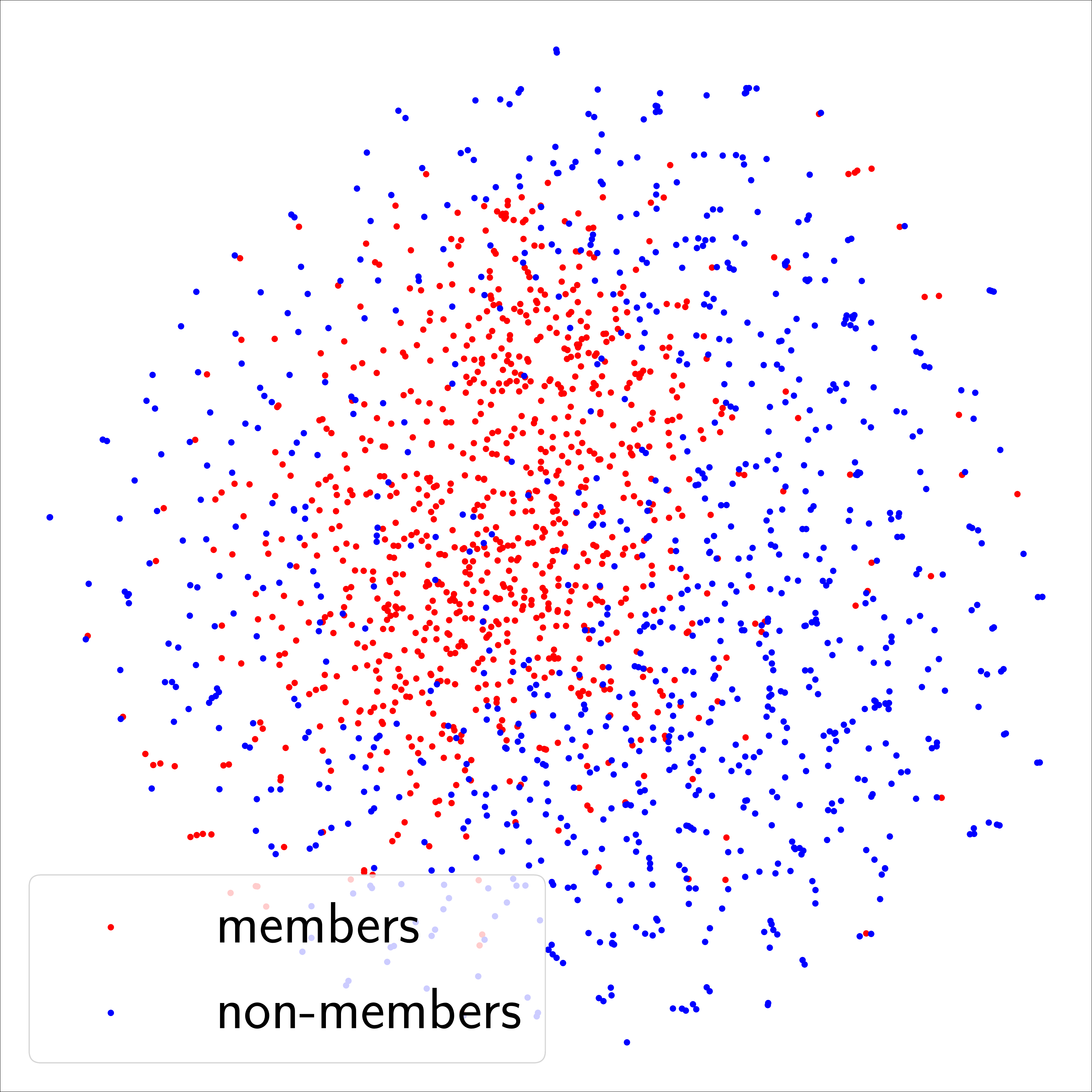}
\includegraphics[width=0.45\columnwidth]{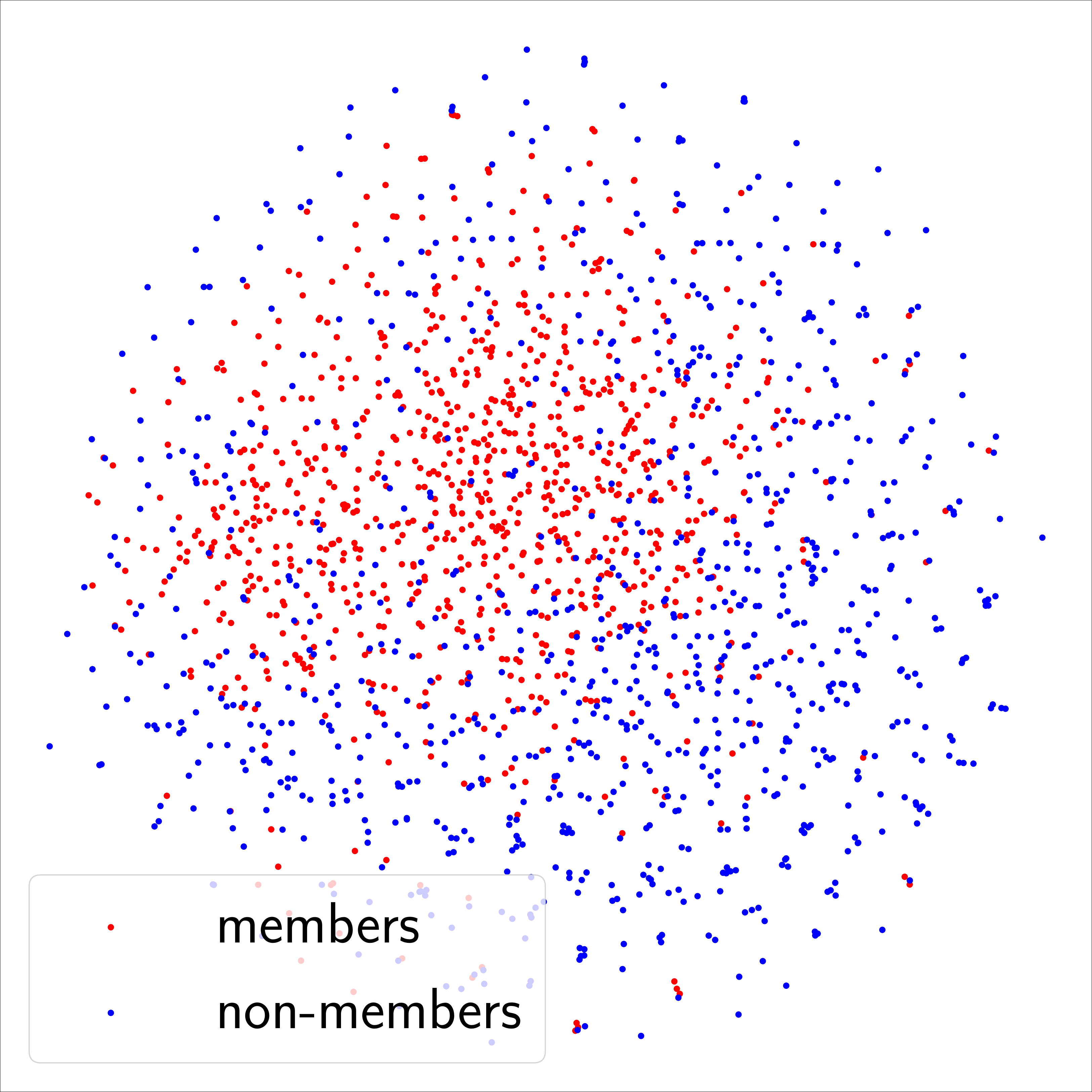}
\caption{lf-2k\_Item\_shadow (left) and lf-2k\_Item\_target (right)}
\label{figure:tsne_LILI}
\end{subfigure}
\begin{subfigure}{\columnwidth}
\centering
\includegraphics[width=0.45\columnwidth]{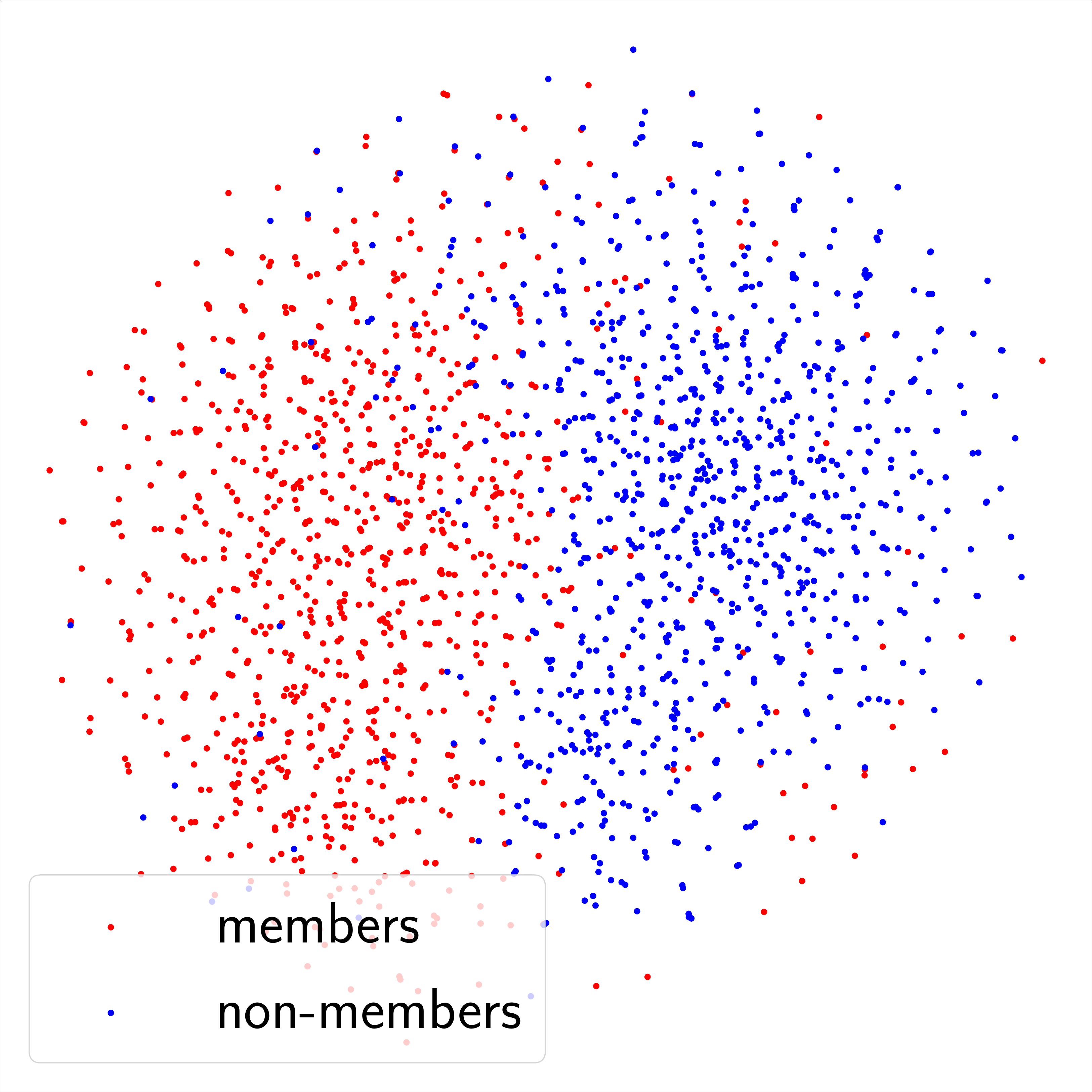}
\includegraphics[width=0.45\columnwidth]{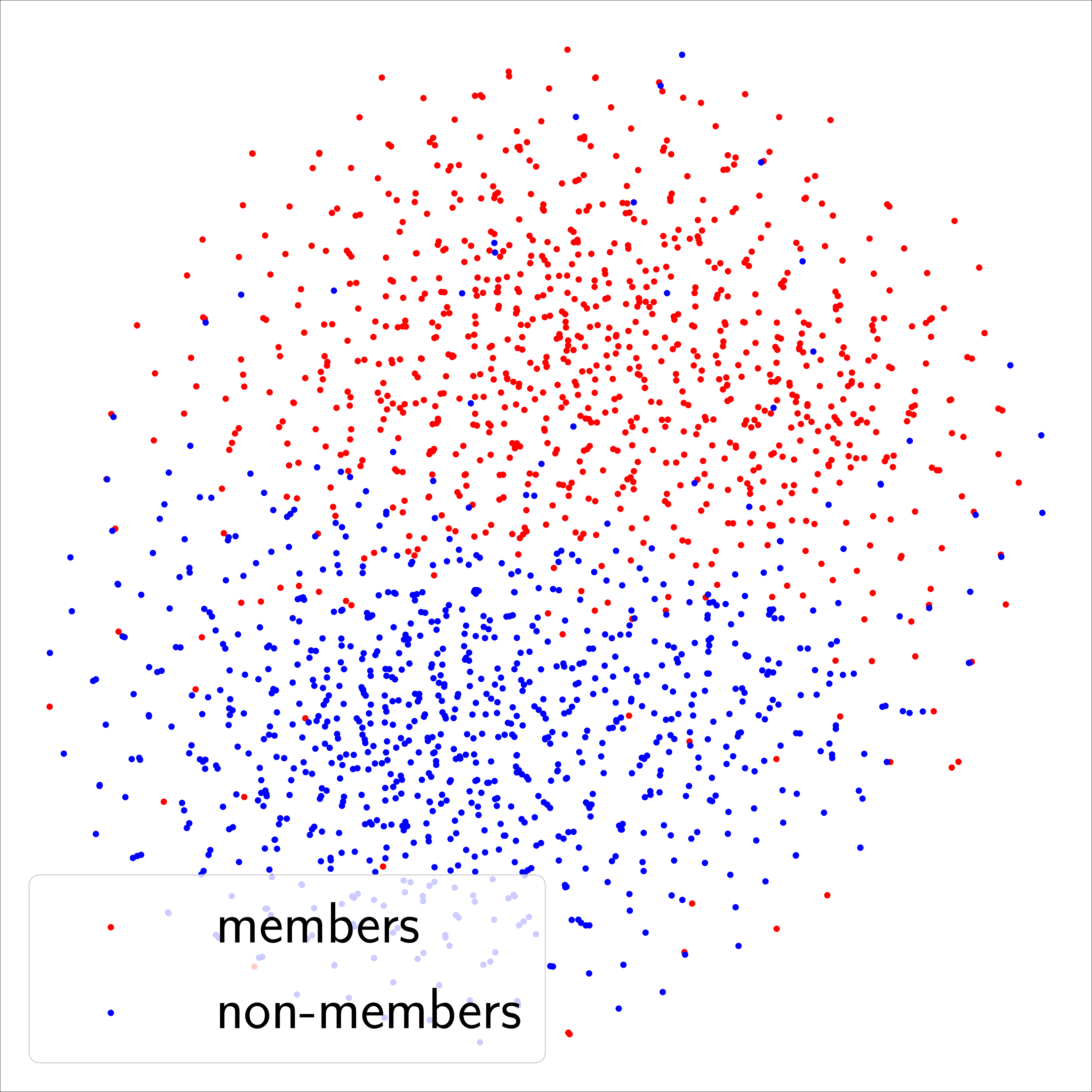}
\caption{lf-2k\_LFM\_shadow (left) and lf-2k\_LFM\_target (right)}
\label{figure:tsne_LLLL}
\end{subfigure}
\caption{Visualization results by t-SNE, where red points denote members and blue points represent non-members.
For the ADM dataset, visualization results, (a) when the shadow and target recommenders are implemented by NCF, and (b) when LFM is used as the shadow and target recommenders, are demonstrated. 
For the lf-2k dataset, visualization results, (c) when the shadow and target recommenders are implemented by Item, and (d) when LFM is adopted as the shadow and target recommenders, are shown.}
\label{figure:tsne_SameAlgorithmSameDataset}
\end{figure*} 

\mypara{Assumption \RNum{1}.}
\begin{table}[!t]
\centering
\caption{The AUC of the attack model against the ADM dataset, under the assumption \RNum{2}.}
\label{table:same_dataset_different_algorithm_amazon}
\setlength{\tabcolsep}{4.3mm}{
\begin{tabular}{l | c   c   c}
\toprule
\makebox[1.5cm][l]{Shadow} & \multicolumn{3}{c}{Target Algorithm} \\
\makebox[1.5cm][l]{Algorithm} & Item & LFM & NCF\\
\midrule
Item & \textbf{0.926} & \textbf{0.885} & 0.750\\
LFM & 0.843 & 0.775 & 0.554\\
NCF & 0.513 & 0.494 & \textbf{0.987}\\
\bottomrule
\end{tabular}
}
\end{table}
\begin{table}[!t]
\centering
\caption{The AUC of the attack model against the lf-2k dataset, under the assumption \RNum{2}.}
\label{table:same_dataset_different_algorithm_lastfm}
\setlength{\tabcolsep}{4.3mm}{
\begin{tabular}{l | c   c   c}
\toprule
\makebox[1.5cm][l]{Shadow} & \multicolumn{3}{c}{Target Algorithm} \\
\makebox[1.5cm][l]{Algorithm} & Item & LFM & NCF\\
\midrule
Item & \textbf{0.939} & 0.796 & 0.793\\
LFM & 0.732 & 0.777 & 0.774\\
NCF & 0.827 & \textbf{0.809} & \textbf{0.916}\\
\bottomrule
\end{tabular}
}
\end{table}
\begin{table}[!t]
\centering
\caption{The AUC of the attack model against the ml-1m dataset, under the assumption \RNum{2}.}
\label{table:same_dataset_different_algorithm_movielens}
\setlength{\tabcolsep}{4.3mm}{
\begin{tabular}{l | c   c   c}
\toprule
\makebox[1.5cm][l]{Shadow} & \multicolumn{3}{c}{Target Algorithm} \\
\makebox[1.5cm][l]{Algorithm} & Item & LFM & NCF\\
\midrule
Item & \textbf{0.998} & 0.792 & 0.706\\
LFM & 0.931 & 0.871 & 0.670\\
NCF & 0.976 & \textbf{0.914} & \textbf{0.998}\\
\bottomrule
\end{tabular}
}
\end{table}
First of all, the target recommender's dataset distribution and algorithm are available.
And, in the paper, these information is the most knowledge that the adversary can gain from the target recommender.
The complete results are shown in~\autoref{figure:SameAlgorithhmSameDataset}, in which we compare our attack with Random Guess.
Then, data points (members and non-members) are visualized in a 2-dimension space by t-distributed Stochastic Neighbor Embedding (t-SNE)~\cite{MH08}.
\autoref{figure:tsne_SameAlgorithmSameDataset} shows the results of the shadow and target distributions for two datasets, where the red points represent members and the blue points represent non-members.
According to the attack and visualization results, we conclude that:
\begin{itemize}
    \item In general, under the assumption that the shadow recommender knows the algorithm and dataset distribution of the target recommender, our attack is very strong.
    There are two main reasons for the effectiveness.
    First, data points of members and non-members are tightly clustered separately.
    Due to the different recommendation methods for members and non-members, generally, the interactions and recommendations of members are more relevant.
    In that case, the intra-cluster distance of members and non-members is much smaller than the inter-cluster distance between them, so that members can be easily distinguished.
    Second, as the adversary has the most knowledge of the target recommender, the shadow recommender can well mimic the target recommender.
    Thus the attack model, which is trained on the ground truth membership generated from the shadow recommender, is able to conduct a membership inference accurately.
    \item When the target recommender uses Item or NCF, our attack performs considerably better on all datasets.
    Specifically, an average AUC of the attack aiming at Item or NCF is 18\% and 20\% higher respectively.
    Compared to the visualization result of LFM, the dissimilarity of the shadow and target distributions of Item or NCF is smaller.
    Thus the attack model can easily deal with the data points which are similar to its training data.
    Besides, our attack performs better when the target dataset is the ml-1m dataset. This is because the user-item matrix of the ml-1m dataset is the densest among all three datasets, which enormously facilitates the item vectorization and attack model training.
\end{itemize}

\begin{figure*}[!t]
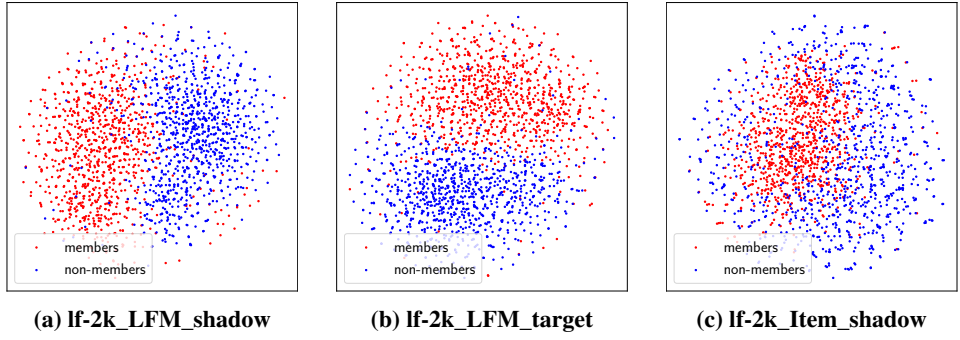

\centering
\begin{subfigure}{0.5\columnwidth}
\centering
\includegraphics[width=0.9\columnwidth]{pic/lf-2k_LFM_shadow.pdf}
\caption{lf-2k\_LFM\_shadow}
\label{figure:lf-2k_LFM_shadow_differentalgorithm}
\end{subfigure}
\begin{subfigure}{0.5\columnwidth}
\centering
\includegraphics[width=0.9\columnwidth]{pic/lf-2k_LFM_target.pdf}
\caption{lf-2k\_LFM\_target}
\label{figure:lf-2k_LFM_target_differentalgorithm}
\end{subfigure}
\begin{subfigure}{0.5\columnwidth}
\centering
\includegraphics[width=0.9\columnwidth]{pic/lf-2k_Item_shadow.pdf}
\caption{lf-2k\_Item\_shadow}
\label{figure:lf-2k_Item_shadow_differentalgorithm}
\end{subfigure}
\caption{Visualization results by t-SNE, where red points are members and blue points are non-members.
For the lf-2k dataset, visualization results, (a) when the shadow recommender is implemented by LFM, (b) when LFM is employed as the target recommender, and (c) when Item is used as the shadow recommender, are demonstrated.}
\label{figure:tsne_differentalgorithm}
\end{figure*}

\mypara{Assumption \RNum{2}.}
To this end, we relax the assumption so that the adversary only has a shadow dataset in the same distribution as the target dataset.
The experimental results are shown in~\autoref{table:same_dataset_different_algorithm_amazon},~\autoref{table:same_dataset_different_algorithm_lastfm} and~\autoref{table:same_dataset_different_algorithm_movielens}, where the attack results of the previous assumption are listed at the diagonals.
Then we depict the visualization results of data points in the shadow and target distributions by t-SNE to show the relationship between members and non-members.
In~\autoref{figure:tsne_differentalgorithm}, the red points are members and the blue points are non-members, which are all from the lf-2k dataset.
We can see from the attack performances as well as the comparisons between the shadow and target distributions that:
\begin{itemize}
    \item When the adversary only gains the knowledge about the target dataset distribution, the attack performances drop as expected but are still strong.
    For instance, on the ADM dataset, when the target recommender uses Item but the shadow recommender uses LFM, the attack performance drops from an AUC of 0.926 to an AUC of 0.843.
    Decreases also appear on the lf-2k dataset and the ml-1m dataset.
    That is to say, even with different recommendation methods, the attack model can still benefit from the similar distributions of the target and shadow datasets to conduct the memebership inference accurately.
    \item An interesting finding is that, the attack on the ml-1m dataset achieves the best overall performance (i.e., 0.873 in terms of average AUC), and the attack performance on the ADM dataset is the worst (i.e., 0.747 in terms of average AUC).
    This is because the user-item matrix built from the ml-1m dataset is the densest while the matrix from the ADM dataset is the sparsest.
    Intuitively, the attack model can learn more information from a denser user-item matrix, leading to a better attack performance.
\end{itemize}
In addition, we do acknowledge that, in some cases, the attack performances are not ideal.
For instance, the attack model against LFM achieves a poor performance (see~\autoref{table:same_dataset_different_algorithm_amazon}). 
Comparing to the other two recommendation algorithms, LFM has higher model complexity, which makes it harder for the adversary to build a similar shadow model.

\begin{figure}[!t]
\centering
\includegraphics[width=0.95\columnwidth]{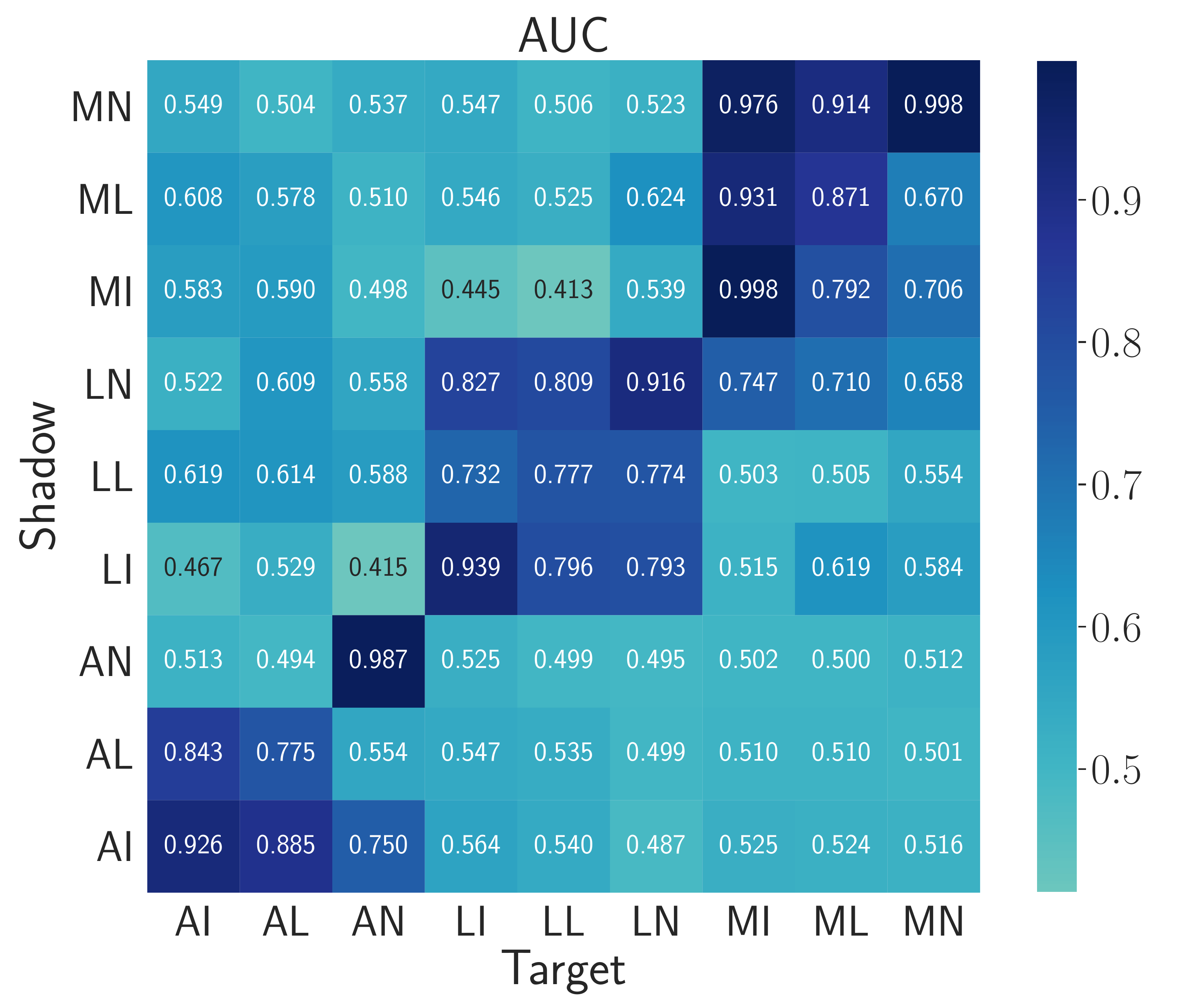}
\caption{The attack performances under the assumption \RNum{3}.
The $x$-axis indicates the target recommender's datasets (the first letter, i.e., ``A,'' ``L'' and ``M'') and algorithms (the second letter, i.e., ``I,'' ``L'' and ``N''), and similarly the $y$-axis represents the shadow recommender's datasets and algorithms.}
\label{figure:heatmap}
\end{figure}

\begin{figure}[!t]
\centering
\begin{subfigure}{\columnwidth}
\centering
\includegraphics[width=0.45\columnwidth]{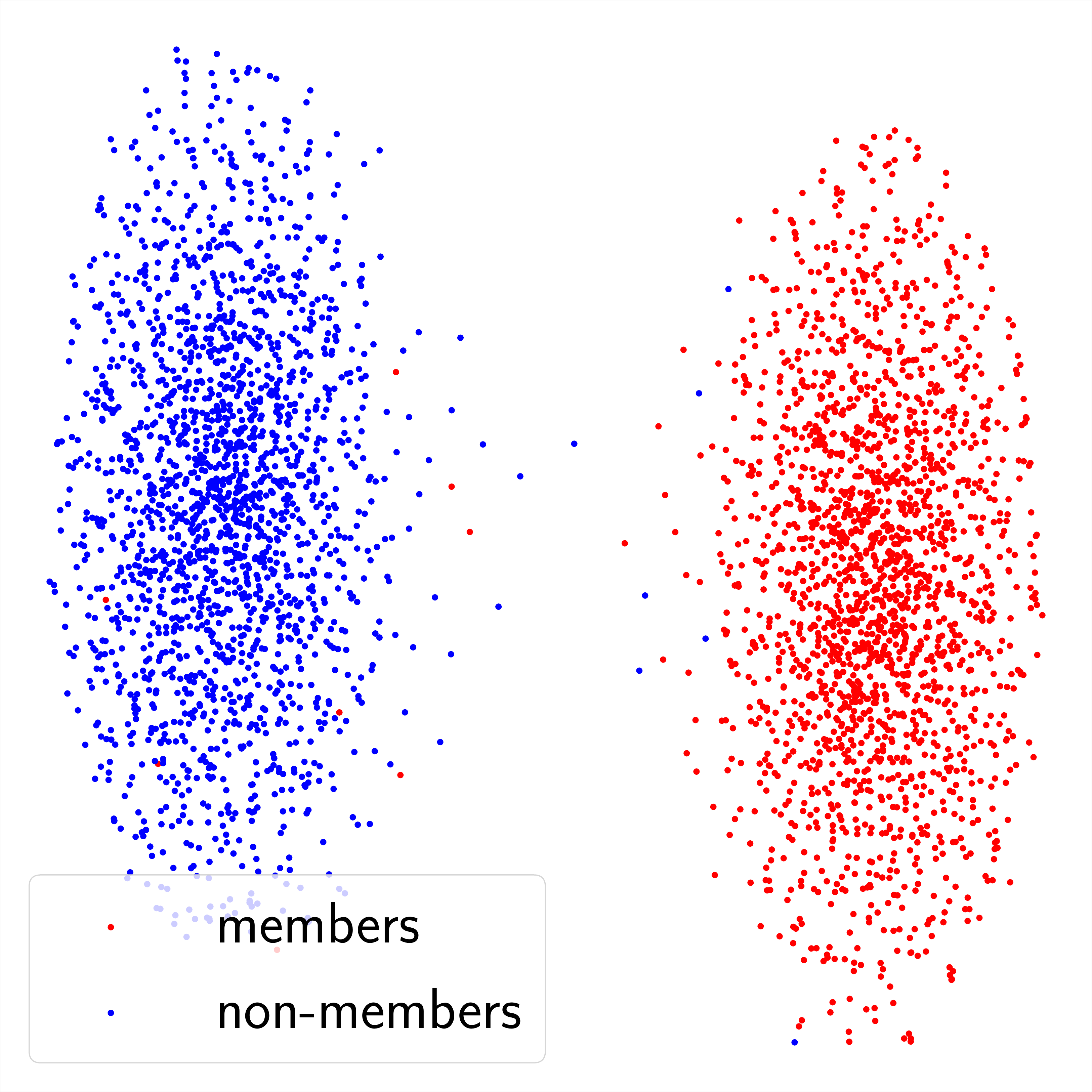}
\includegraphics[width=0.45\columnwidth]{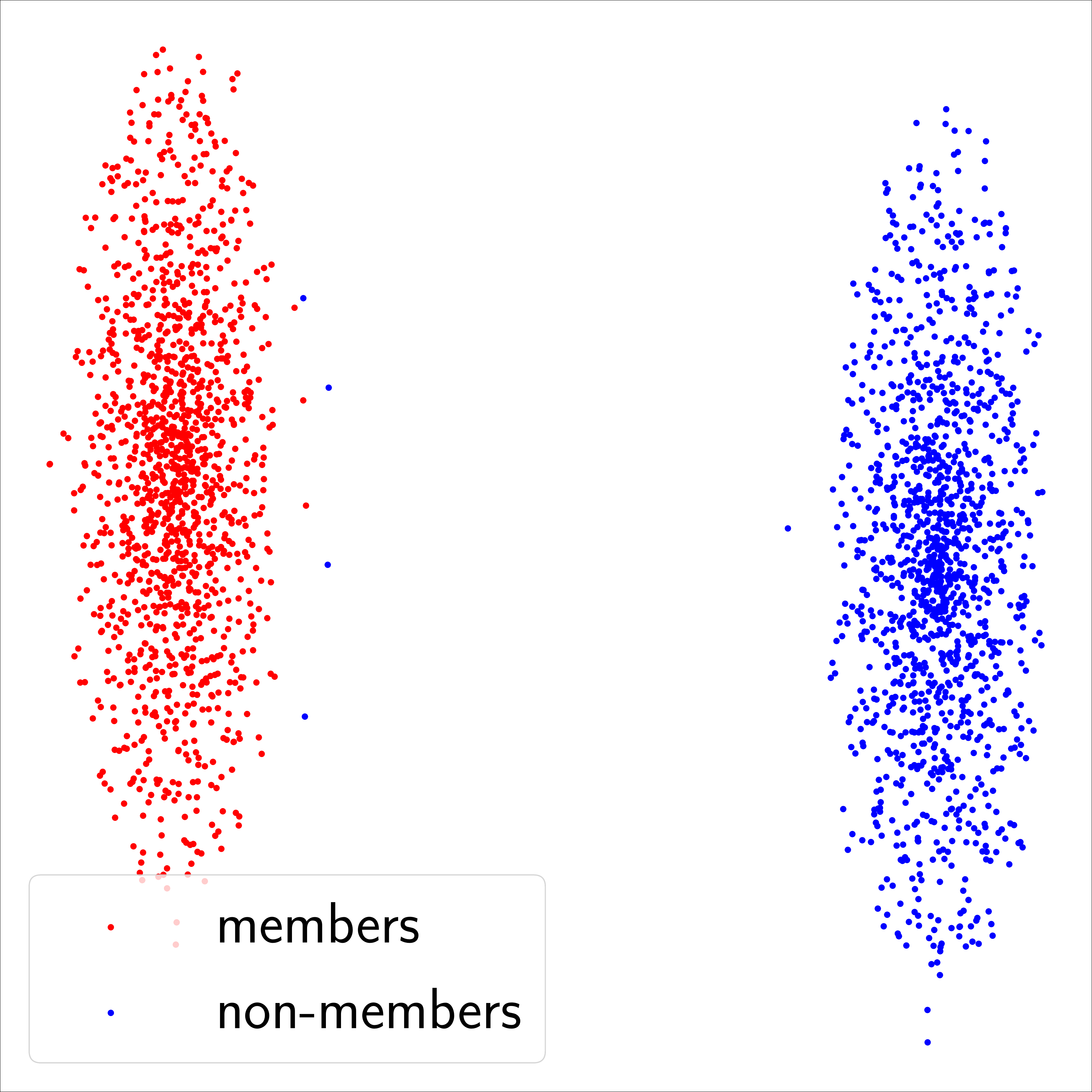}
\caption{ml-1m\_LFM\_shadow (left) and ADM\_Item\_target (right)}
\label{figure:MLAI}
\end{subfigure}
\begin{subfigure}{\columnwidth}
\centering
\includegraphics[width=0.45\columnwidth]{pic/ADM_LFM_shadow.pdf}
\includegraphics[width=0.45\columnwidth]{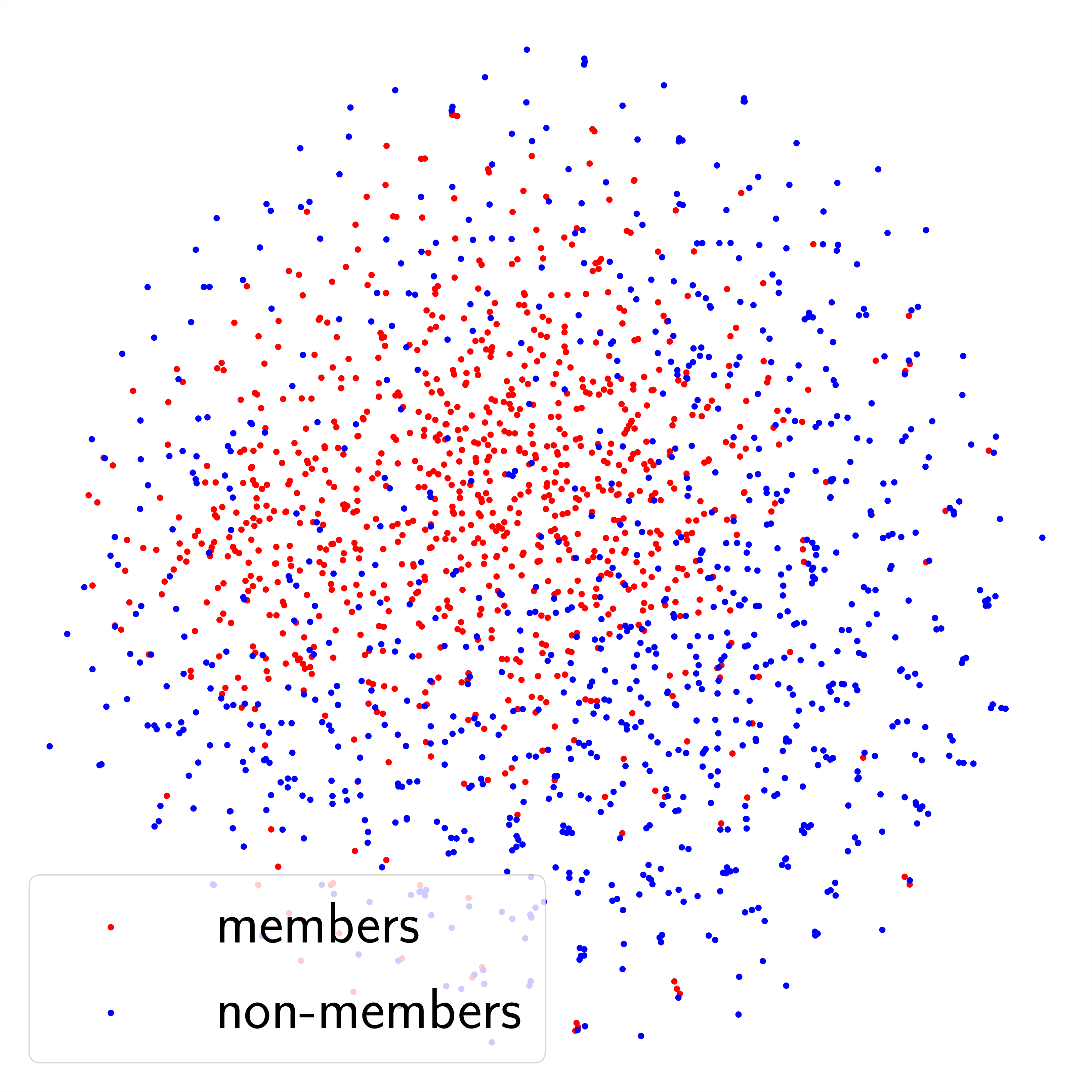}
\caption{ADM\_LFM\_shadow (left) and lf-2k\_Item\_target (right)}
\label{figure:ALLI}
\end{subfigure}
\caption{The visualization results of MLAI v.s. ALLI}
\label{figure:specialcase_assumption3}
\end{figure}

\mypara{Assumption \RNum{3}.}
Finally, we further conduct evaluations when the adversary neither has a shadow dataset in the same distribution as the target dataset nor knows the target algorithm.
All experimental results are shown in~\autoref{figure:heatmap}.
Note that, the attack results of the assumption \RNum{1} are listed at the back-diagonal and the attack performances of the assumption \RNum{2} are shown in the three $3 \times 3$ block back-diagonal matrices.
Analysing the results, we draw conclusions that:
\begin{itemize}
    \item Even under the minimum assumption, our attack can still achieve strong performances in most cases.
    For instance, when the target recommender is established by LFM on the ml-1m dataset and the adversary uses NCF to build a shadow recommender on the lf-2k dataset, our attack achieves an AUC of 0.710.
    \item In some cases, when the adversary knows less information about the target recommender, the attack even achieves better performances.
    For instance, when the adversary builds a shadow recommender by NCF on the lf-2k dataset to mimic the target recommender which uses Item on the ml-1m dataset, our attack achieves an AUC of 0.747.
    Meanwhile, with the knowledge that the target dataset is the ml-1m dataset, the adversary uses LFM to establish a shadow recommender when the target recommender uses NCF, our attack only achieves an AUC of 0.670.
    To explain this, we adopt the t-SNE algorithm to visualize user feature vectors for the ``MLAI'' and ``ALLI'' attacks.
    The visualization results in~\autoref{figure:specialcase_assumption3} show that the distributions of feature vectors generated by the shadow model ``ML'' and target model ``AL'' are more similar than the distributions generated by ``Al'' and ``LI''.
    Therefore, the ``MLAI'' attack performs better than the ``ALLI'' attack.
\end{itemize}
In summary, our attack can effectively conduct a membership inference against recommender systems, even with the limited knowledge.

\subsection{Hyperparameters}
\label{section:hyperparameters}

\begin{figure*}[!t]
\centering
\begin{subfigure}{0.62\columnwidth}
\centering
\includegraphics[width=\columnwidth]{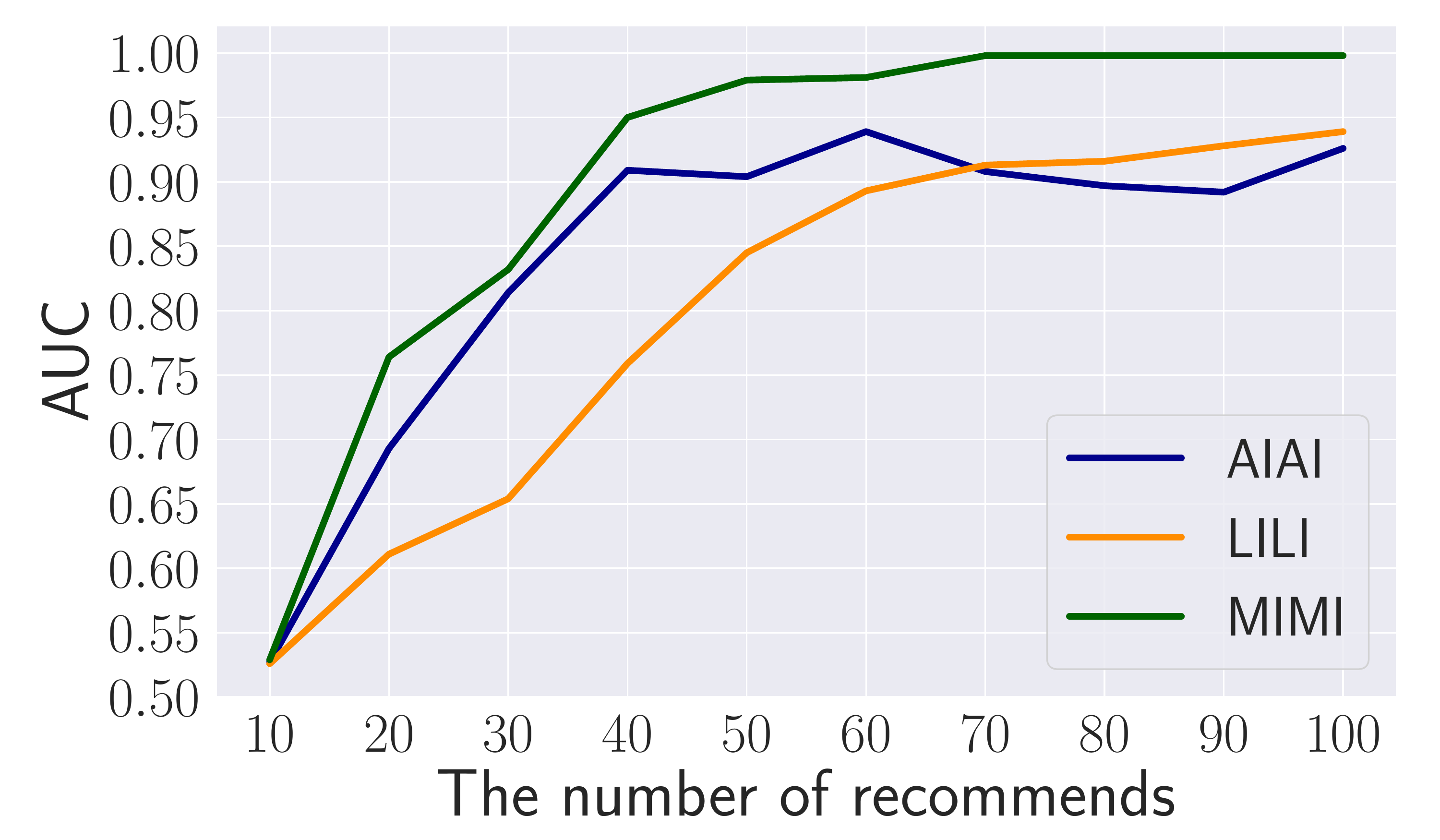}
\caption{The attack performances against the number of recommendations $k$.}
\label{figure:k}
\end{subfigure}
\hspace{2mm}
\begin{subfigure}{0.62\columnwidth}
\centering
\includegraphics[width=\columnwidth]{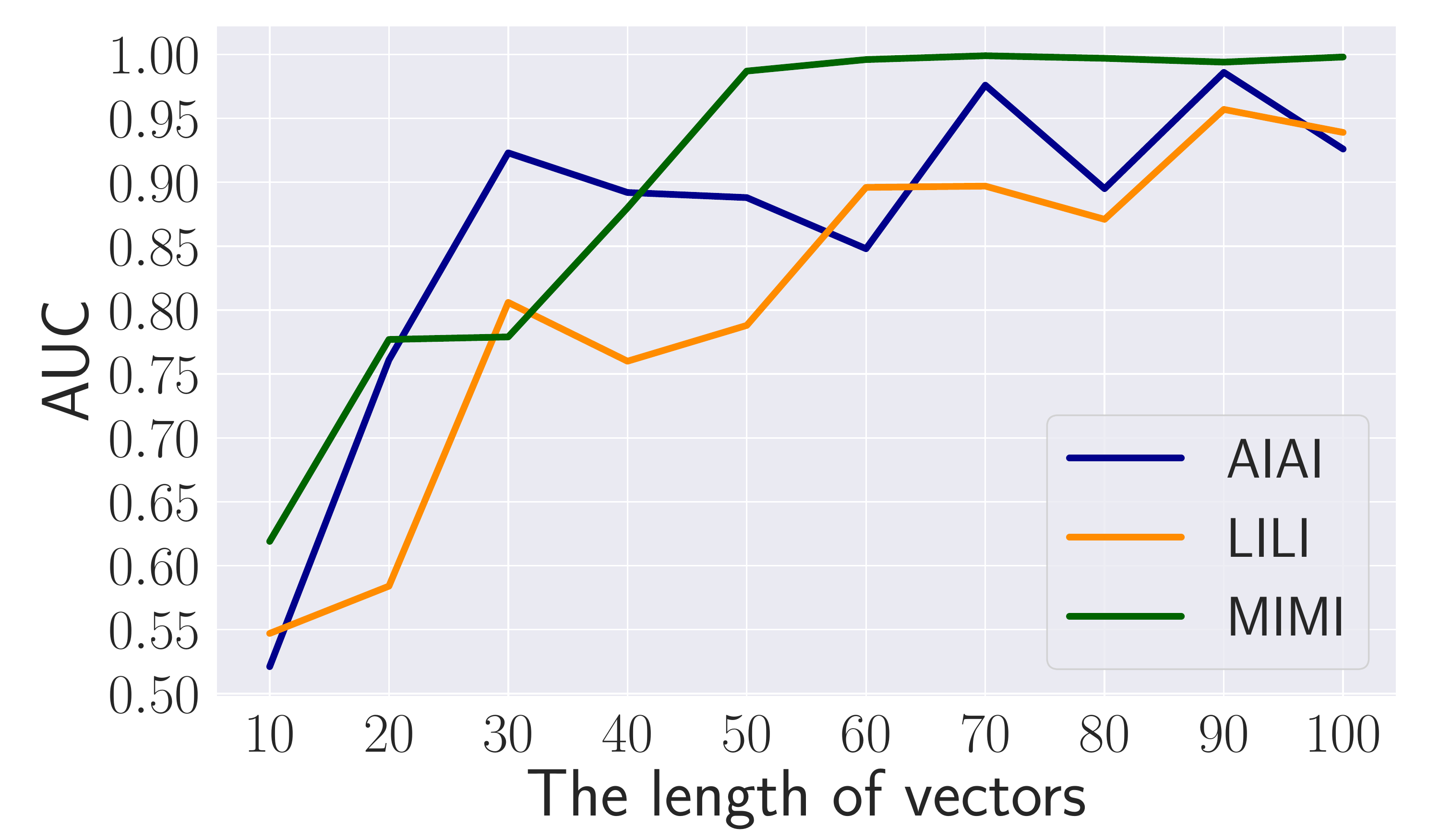}
\caption{The attack performances against the length of vectors $l$.}
\label{figure:l}
\end{subfigure}
\hspace{2mm}
\begin{subfigure}{0.62\columnwidth}
\centering
\includegraphics[width=\columnwidth]{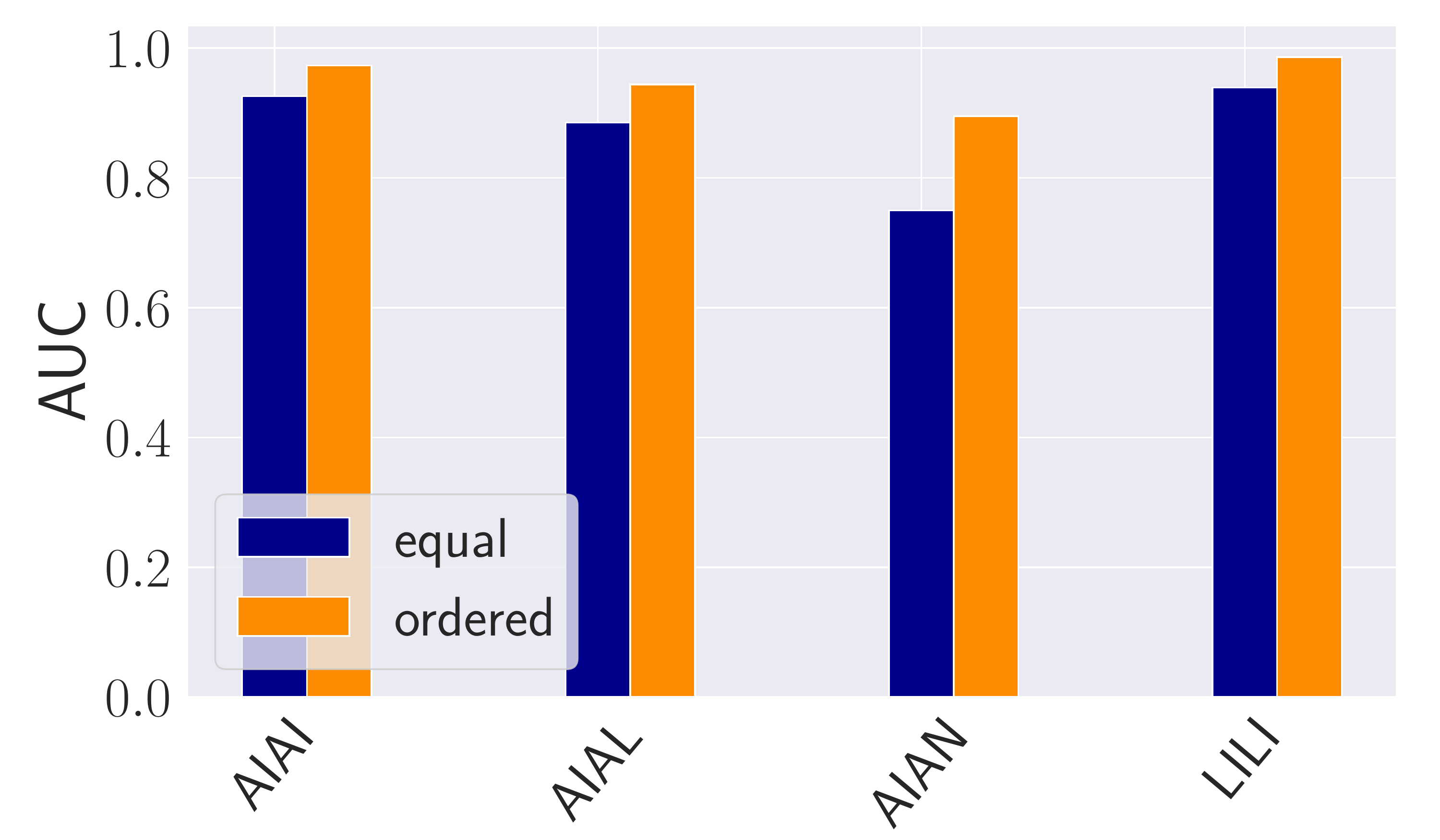}
\caption{The attack performances against the weights of recommendations.}
\label{figure:weights}
\end{subfigure}
\caption{The attack performances of analysing the influences of hyperparameters.}
\label{figure:hyperparameters}
\end{figure*} 

\begin{figure*}[!t]
\centering
\begin{subfigure}{0.9\columnwidth}
\centering
\includegraphics[width=\columnwidth]{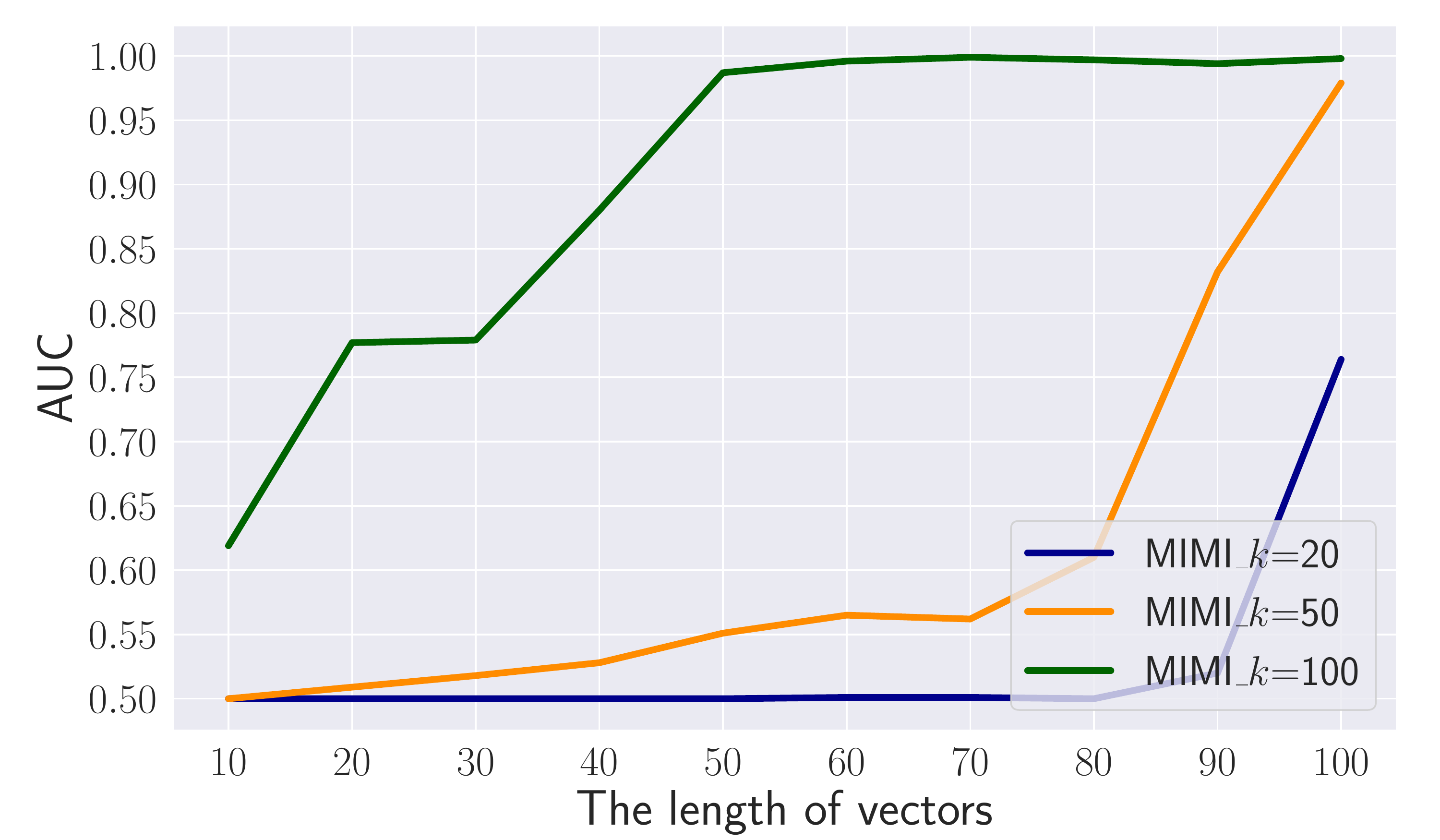}
\caption{The attack performances with different $l$, when $k=20$, $k=50$ and $k=100$.}
\label{figure:K_20_50_100}
\end{subfigure}
\hspace{2mm}
\begin{subfigure}{0.9\columnwidth}
\centering
\includegraphics[width=\columnwidth]{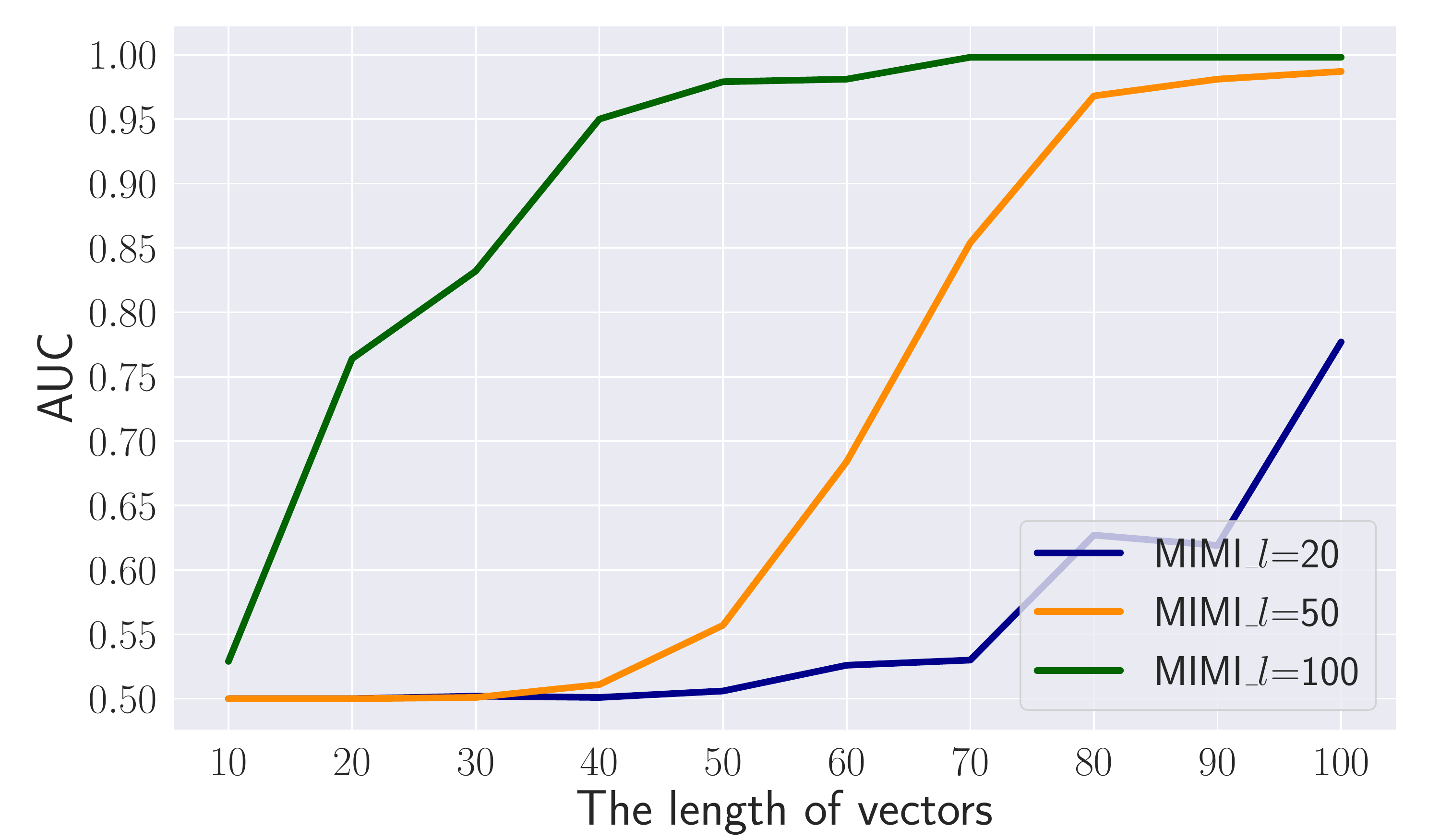}
\caption{The attack performances with different $k$, when $l=20$, $l=50$ and $l=100$.}
\label{figure:L_20_50_100}
\end{subfigure}
\caption{The attack performances to investigate ``which is more important, $k$ or $l$?''}
\label{figure:KvsL}
\end{figure*}

In this section, we analyse the influences of hyperparameters, including the number of recommendations $k$, the length of vectors $l$ and the weights of recommendations.
\autoref{figure:hyperparameters} shows the experimental results.

\mypara{The Number of Recommendations $k$.}
We evaluate our experiments with different values of $k$ from 10 to 100, in order to explore the influence of $k$ on the attack.
\autoref{figure:k} shows the attack performance against the number of recommendations.
When the number of recommendations is less than 50, the attack performance improves with the increase of $k$.
Then the performance maintains stable when $k$ goes beyond 50.
These results show that the attack model gains more information when the number of recommendations increases.
However, the attack model cannot gain more information infinitely when the number of recommendations is large enough.

\mypara{The Length of Vectors $l$.}
We evaluate our experiments with different values of $l$ from 10 to 100, in order to explore the influence of $l$ on the attack.
\autoref{figure:l} shows the attack performance against the length of vectors.
Similar to~\autoref{figure:k}, when the length of vectors is less than 50, the attack performance improves with the increase of $l$.
Then, in general, no obvious improvement of the performance is observed when $l$ goes beyond 50.
These results show that the representation power of the attack model becomes stronger, as a larger length of vectors can provide more dimensional perspectives.
However, the attack model cannot improve its representation power infinitely when the length of vectors is large enough.

\mypara{The Weights of Recommendations.}
We evaluate our experiments with different designs for the weights of recommendations.
In the real world, the items recommended to a user are provided in the form of an ordered sequence.
And, compared to the items at the back of the sequence, the ones in front of the sequence are more likely to be preferred by the user.
Thus we evaluate two methods of assigning weights to items at different positions in the sequence.
One is that all items are assigned the same weight of $\frac{1}{k}$.
And the other is to assign a weight of $\frac{k-i+1}{\sum_{n=1}^{k}n}$ to the $i^{th}$ item in the sequence.
Same as mentioned above, we denote $k$ as the number of recommendations.
As shown in~\autoref{figure:weights}, we find that considering the order of recommendations can obviously promote attack performances.

\subsection{Extensive Analysis}
\label{section:extensiveanalysis}

In this section, we study five interesting questions and give further results in order to comprehensively investigate our attack method.

\mypara{Which Is More Important, the Length of Feature Vector $l$ or the Number of Recommendations $k$?}
In addition to the analyses about hyperparameters in~\autoref{section:hyperparameters}, we also investigate ``which is more important, $k$ and $l$?''.
Two more experiments are conducted on the ``MIMI'' attack, with different $l$ when $k$ is set to 20, 50 and 100 (see~\autoref{figure:K_20_50_100}), and with different $k$ when $l$ is set to 20, 50 and 100 (see~\autoref{figure:L_20_50_100}).
As the results show, both the number of recommendations ($k$) and the length of vectors ($l$) influence attack performances substantially. Specifically, when $k$ reduces from 100 to 20, the AUC score drops from 0.998 to 0.764.
Similarly, as $l$ reduces from 100 to 20, the AUC score descends from 0.998 to 0.817.

\mypara{What Is the Impact of the Dataset Size?}
Considering the size of the training dataset imposes huge impacts on machine learning models, we conduct evaluations regarding the size of the shadow dataset.
Specifically, the size of the shadow dataset is reduced to 90\%, 80\%, and 70\% of the original size.
Note that, the ratio of members to non-members keep unchanged for the dataset balance.
For the ``LLLL'' attack, the AUC scores of the attack performances are decreased to 0.633, 0.714, and 0.746, respectively, when the size of the shadow dataset is 70\%, 80\%, and 90\% of the original size. Comparing to the original AUC of 0.777, we can conclude that a larger shadow dataset usually leads to a better-trained attack model.

\mypara{Why Use an MLP as the Attack Model?}
\begin{table}[!t]
\centering
\caption{The AUC of the K-Means algorithm on the lf-2k dataset.}
\label{table:k-means_lf-2k}
\setlength{\tabcolsep}{4.3mm}{
\begin{tabular}{l | c   c   c}
\toprule
\makebox[1.5cm][c]{} & \multicolumn{3}{c}{Target Algorithm} \\
\makebox[1.5cm][c]{} & Item & LFM & NCF\\
\midrule
Item & \textbf{0.939} & 0.796 & 0.793\\
LFM & 0.732 & 0.777 & 0.774\\
NCF & 0.827 & \textbf{0.809} & \textbf{0.916}\\
\midrule
K-Means & \emph{0.649} & \emph{0.717} & \emph{0.730}\\
\bottomrule
\end{tabular}
}
\end{table}
\begin{table}[!t]
\centering
\caption{The AUC of the K-Means algorithm on the ml-1m dataset.}
\label{table:k-means_ml-1m}
\setlength{\tabcolsep}{4.3mm}{
\begin{tabular}{l | c   c   c}
\toprule
\makebox[1.5cm][c]{} & \multicolumn{3}{c}{Target Algorithm} \\
\makebox[1.5cm][c]{} & Item & LFM & NCF\\
\midrule
Item & \textbf{0.998} & 0.792 & 0.706\\
LFM & 0.931 & 0.871 & 0.670\\
NCF & 0.976 & \textbf{0.914} & \textbf{0.998}\\
\midrule
K-Means & \emph{0.805} & \emph{0.720} & \emph{0.719}\\
\bottomrule
\end{tabular}
}
\end{table}
To demonstrate the effectiveness of our attack model, we evaluate the attacks utilizing K-Means to distinguish non-members from members on the lf-2k and ml-1m datasets.
The results are shown in~\autoref{table:k-means_lf-2k} and~\autoref{table:k-means_ml-1m}, where Item, LFM, and NCF in the first column are shadow algorithms for our attack, and K-Means is used to cluster non-members and members.
Since there are only two classes, i.e., members and non-members, the number of classes $K$ for K-Means is set to 2.
From the results, we can conclude that our attack outperforms K-Means largely, indicating the validity of our attack model.
For instance, when the K-Means algorithm infers the membership status from the target recommender ``LI'', the performance is much worse than our attack (0.649 v.s. 0.939).

\begin{figure}[!t]
\centering
\includegraphics[width=0.95\columnwidth]{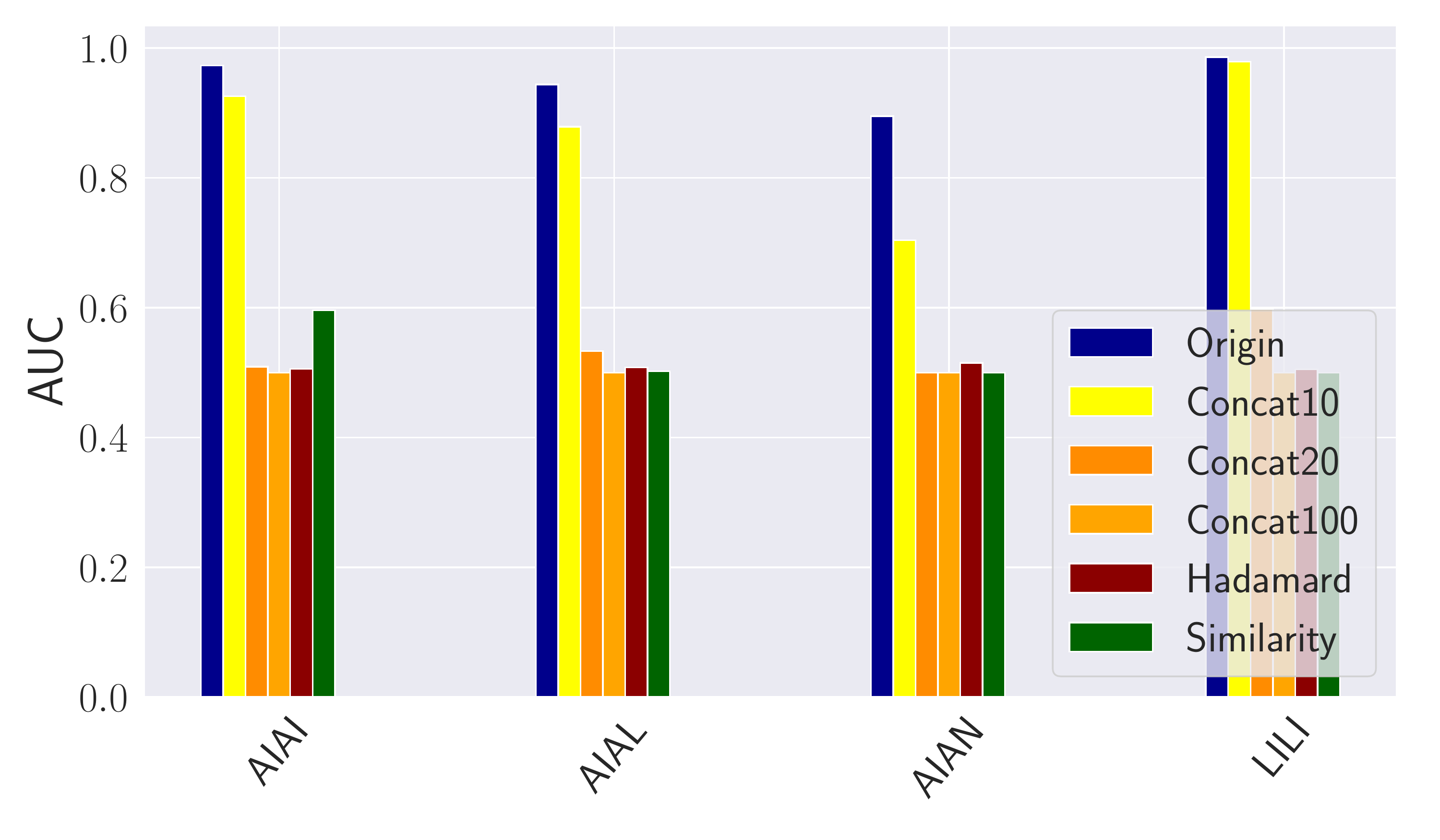}
\caption{The attack performances with different user feature generation methods}
\label{figure:labelgeneration}
\end{figure}
\mypara{Which Is the Best User Feature Generation Method?}
To further verify the effectiveness of our attack, we adopt different aggregation methods to generate user feature vectors.
Besides the method used in our attack, \textbf{Origin}, we evaluate 5 user feature generation methods.
\begin{itemize}
    \item \textbf{Concat10} concatenates feature vectors of the first 10 interactions and the first 10 recommendations for each user.
    \item \textbf{Concat20} concatenates feature vectors of the first 20 interactions and the first 20 recommendations for each user.
    \item \textbf{Concat100} concatenates feature vectors of the first 20 interactions and all the recommendations (i.e., $k=100$) for each user.\footnote{As the number of interactions for different users are different, using all interactions for concatenation will result in different lengths of user feature vectors.
    Therefore, we only take the first 20 interactions into consideration since the least number of interactions in datasets is 20.}
    \item \textbf{Hadamard} respectively conducts Hadamard products on feature vectors of all the interactions and recommendations to obtain two vectors for each user.
    Afterwards, following the similar steps in~\autoref{section:MIA}, we use the difference of these two vectors as the user feature vector.
    \item \textbf{Similarity} first conducts dot products between feature vectors of each recommendation and all interactions, then concatenates the average of dot product results for each recommendation into the user feature vector.
\end{itemize}

The results are shown in~\autoref{figure:labelgeneration}.
We find that our method outperforms all the other aggregation methods.
For instance, on the settings of AIAI, our method outperforms Concat10 and Similarity by 5\% and 63\%.
These results demonstrate the effectiveness of our user feature generation method in the attack process.

\begin{figure}[!t]
\centering
\includegraphics[width=0.95\columnwidth]{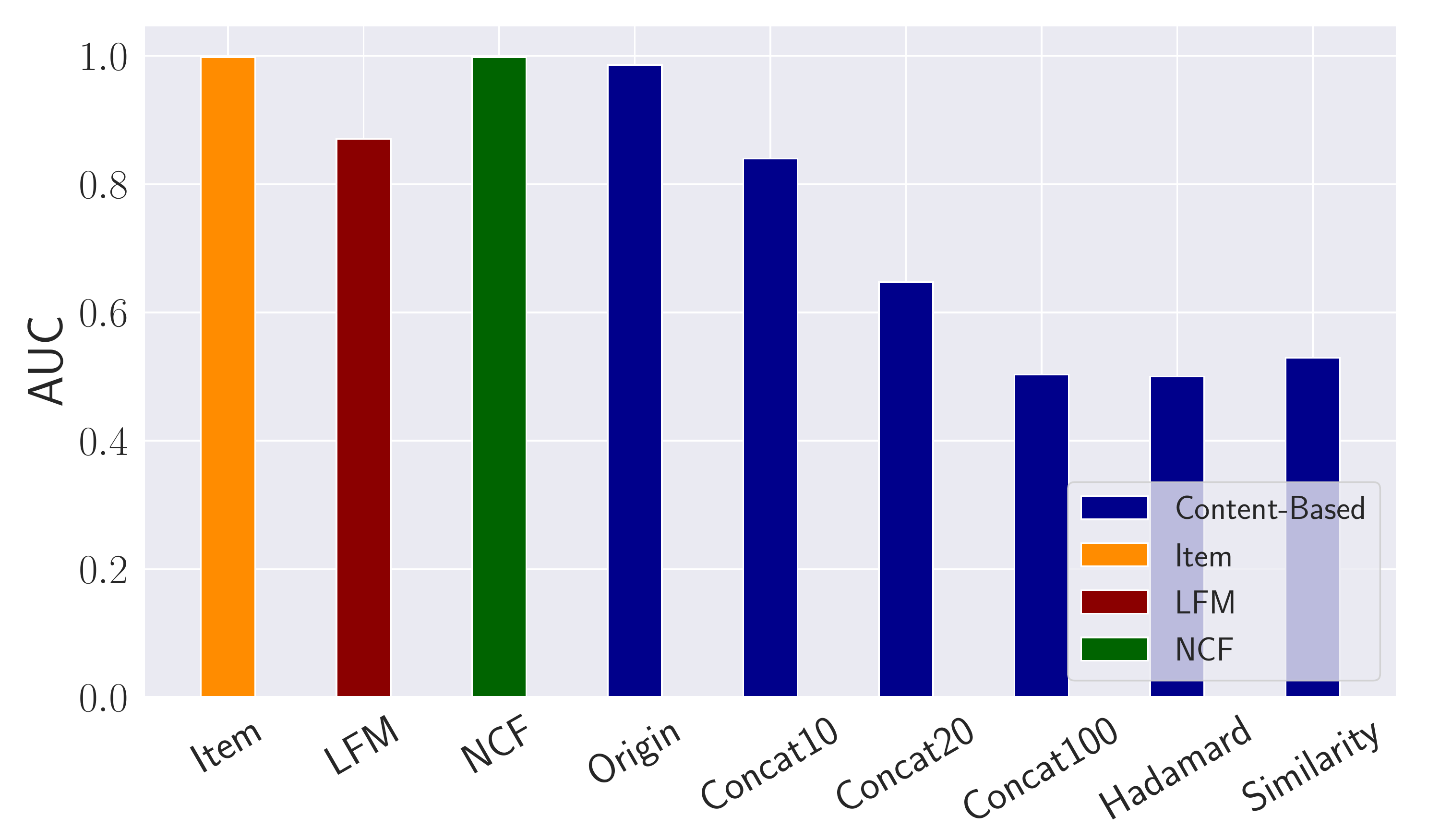}
\caption{The attack performances for the content-based recommender system on the ml-1m dataset.}
\label{figure:content}
\end{figure}
\mypara{Is It Possible to Generalize Our Attacks to Content-Based Recommender Systems?}
To further verify the effectiveness of our attack, we conduct evaluations on the membership inference attacks against a content-based recommender system.
Different from the recommendation algorithms used in the previous attacks, a content-based recommender system aims to distinguish users' likes from dislikes based on their \emph{metadata} (such as description of items and profiles of users)~\cite{WF20}.
The evaluation is conducted on the ml-1m dataset with information about items and users, under the assumption \RNum{1}, i.e., the target recommender's algorithm and dataset distribution are available.

The results are depicted in~\autoref{figure:content}. 
We conclude that our attack achieves a strong performance against a content-based recommendation algorithm (i.e., 0.986 in terms of AUC), indicating a high generalization ability of our attack model.
Furthermore, we evaluate this attack using different feature generation methods mentioned above.
Results show that our aggregation method also outperforms other baselines on a content-based recommendation system.

\subsection{Summary}

The experimental results show that our attack can conduct an effective membership inference against recommender systems.
When the adversary knows the algorithm and dataset distribution of the target recommender, the attack achieves the strongest performance.
Later in the experiments, we gradually relax the assumptions and show that our attack can still effectively conduct the membership inference, demonstrating that our attack has a good generalization ability.

Furthermore, we explore the influences of hyperparameters.
With the increase of the number of recommendations $k$ and the length of vectors $l$, the attack performance improves or maintains stable, however, the cost continuously increases.
In that case, we are able to find a balance, where the attack performance is strong and the cost is affordable.
And the exploration of the weights of recommendations shows that the more information is available, the more powerful the attack is.

\section{Defense}
\label{section:defense}

The above experiments show the effectiveness of our attack. 
Meanwhile, to defend the membership inference against recommender systems, we also propose a countermeasure, named Popularity Randomization, to mitigate the attack risk. 
In the original setting in~\autoref{section:MIA}, non-members are provided with the most popular items. As a result, feature vectors of non-members are extremely similar and easily distinguished from members.
To address this problem, we increase the randomness of non-members' recommendations. Specifically, we first select candidates from the most popular items.
Then, a random selection is conducted on the candidates, i.e., we randomly pick 10\% of candidates as recommendations for non-members. 
The detailed methodology is demonstrated in Appendix~\ref{section:defensemethodology}.

\begin{figure}[!t]
\centering
\includegraphics[width=0.95\columnwidth]{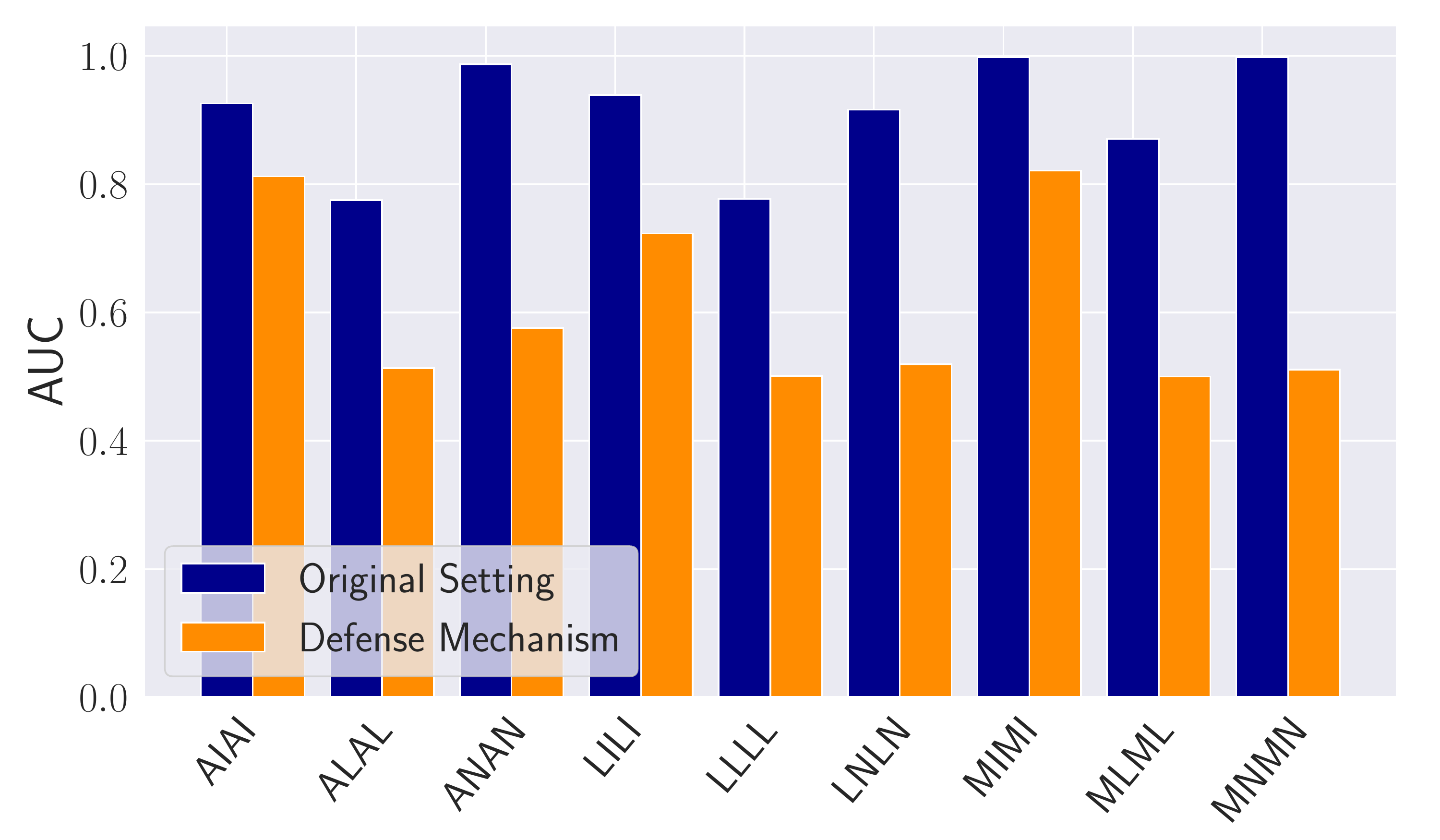}
\caption{Comparisons of attack performances before and after deploying the defense mechanism.}
\label{figure:defense}
\end{figure}

To evaluate the effectiveness of the defense mechanism, we conduct experiments under the assumption that the dataset distribution and algorithm of the target recommender are available.
\autoref{figure:defense} shows the attack performances before and after deploying the defense mechanism.
The blue bar denotes the attack performances with the original setting, i.e., the popularity recommendation algorithm.
And the orange bar represents the attack performances with the defense mechanism, i.e., Popularity Randomization.
From the results, we conclude that Popularity Randomization considerably decreases the performance of our attack.
Specifically, the defense mechanism decreases the AUC scores of the attack model by more than $12\%$, $33\%$ and $41\%$ respectively when the target recommender uses Item, LFM or NCF.
With the defense strategy, attacking the target recommender using LFM achieves the lowest AUC score on all three datasets.
When the target recommender using LFM, our attack with the defense mechanism only achieves 0.513, 0.501, and 0.500 AUC scores on the ADM dataset, the lf-2k dataset and the ml-1m dataset, respectively (detailed explanations are demonstrated in Appendix~\ref{section:defenseresults}).
Besides, as~\autoref{figure:defense} shows, the attack performances against NCF decrease most hugely.
For instance, the AUC score of the attack against NCF on the ADM dataset drops from 0.987 to 0.576.
In contrast, the attack with Popularity Randomization against Item can still achieve strong performances.
For instance, on the ADM dataset, the attack with Popularity Randomization can attain 0.812 in terms of AUC when the target algorithm is Item.
Compared to the attack with the original setting, the defense mechanism only achieves a 12\% drop in the attack performance.
Item is the simplest one among the three recommendation methods, which makes it easier for the adversary to build a similar shadow recommender with the target recommender.
This leads to a stronger attack but more ineffective defense.
In contrast, the other two recommender systems have more complex model structures, leading to substantial decreases in attack performances with the defense strategy.

In addition, visualization results and impacts on recommendation performances are comprehensively analyzed in Appendix~\ref{section:defenseresults} and Appendix~\ref{section: Impacts on Recommendation Performances}, respectively.

\section{Related Work}
\label{section:relatedwork}

\mypara{Membership Inference.}
The goal of membership inference is to infer whether a target data sample is used to train a machine learning model~\cite{SSSS17,YGFJ18,SZHBFB19,LZ21,LF20,CTWJHLRBSEOR20,NSH18,JSBZG19,SH21}.
Shokri et al.~\cite{SSSS17} propose the first membership inference attack in this domain.
The authors have made several key assumptions for the adversary, such as multiple shadow models and a shadow dataset which comes from the same distribution as the target model's training datasset.
Salem et al.~\cite{SZHBFB19} gradually relax these assumptions and broaden the scenarios of membership inference attacks.
Later, Nasr et al.~\cite{NSH19} conduct a comprehensive membership privacy assessment in both centralized and federated learning setting.
In particular, they propose the first membership inference when the adversary has white-box to the target model.
Other research has shown that membership inference is effective under other machine learning settings, such as generative models~\cite{HMDC19}, federated learning~\cite{MSCS19,CYZF20}, and natural language models~\cite{SS19}.
Besides, a plethora of other attacks have been proposed against machine learning models~\cite{BCMNSLGR13,TKPGBM17,JOBLNL18,CW17,PMSW18,CLEKS19,JCBKP20,TZJRR16,GDG17}.

\mypara{Item-Based Recommendation Algorithms.}
Item-based recommendation techniques have been applied in various scenarios~\cite{SKKR01, DK04, K01}.
Sarwar et al.~\cite{SKKR01} explore item-based collaborative filtering (CF) techniques which enhance the scalability and quality of the CF-based algorithms.
Besides, Deshpande and Karypis~\cite{DK04,K01} present item-based top-$N$ recommendation algorithms to promote the efficiency and performance.

\mypara{Latent Factor Models.}
LFM aims to find some latent factors and is commonly implemented by Matrix Factorization (MF)~\cite{PD05,SM07,K08,K09}.
Polat et al.~\cite{PD05} combine SVD-based Collaborative Filtering with privacy to achieve accurate predictions while preserving privacy.
Later, Salakhutdinov et al.~\cite{SM07} propose the Probabilistic Matrix Factorization which scales linearly with the number of observations and performs well on very sparse and imbalanced datasets.
Koren~\cite{K08} presents an integrated model that combines the neighborhood and LFM, which optimizes a global cost function and integrates implicit feedback into the model.
Furthermore, Koren~\cite{K09} presents a methodology for modeling time drifting user preference in the context of recommender systems.

\mypara{Neural Collaborative Filtering.}
With the advancement of deep learning techniques, recommendation algorithms with neural networks has been in blossom~\cite{HLZNHC17, BWZZ17, CCCR19, SWW18, SWZFHW20}.
He et al.~\cite{HLZNHC17} propose the first framework for collaborative filtering based on neural networks to model latent features of users and items.
They show that MF can be interpreted as a specialization of NCF and utilize a MLP to endow NCF modelling with a high level of non-linearities.
Later, Bai et al.~\cite{BWZZ17} present a model which integrates neighborhood information into NCF, namely Neighborhood-based Neural Collaborative Filtering.
Another recent work is that Chen et al.~\cite{CCCR19} design a Joint Neural Collaborative Filtering model which enables deep feature learning and deep user-item interaction modeling to be tightly coupled and jointly optimized in a single neural network.

\section{Discussion}
\label{section:discussion}

In the previous evaluations, our attack shows its effectiveness as well as strong generalization ability.
Moreover, the proposed defense mechanism, Popularity Randomization, can also mitigate the attack performances considerably.
Furthermore, to obtain a comprehensive understanding of membership inference attacks, in this section, we focus on three important factors that largely influence attack performances: the choice of datasets, the selection of recommendation algorithms, and distributions of generated user features.
The detailed explanations are demonstrated as follows:

\mypara{The Choice of Datasets.}
The dataset with a denser user-item matrix leads to better attack performances.
The richer information in a denser user-item matrix considerably facilitates the process of the item vectorization and attack model training.
From the results and analyses under the assumption \RNum{2} in~\autoref{section:attackperformance}, we can see that the attack on the ml-1m dataset achieves the best overall performance (i.e., 0.873 in terms of average AUC) as the user-item matrix built from this dataset is the densest.

\mypara{The Selection of Recommendation Algorithms.}
It is easier for the adversary to attack against a recommender system with a simpler model structure.
As the results (see~\autoref{table:same_dataset_different_algorithm_amazon}) under the assumption \RNum{2} show, the attack against LFM achieves poor performances. 
Comparing to the other two recommendation algorithms, LFM has higher model complexity, which makes it harder to attack.
Meanwhile, defending a simpler recommender system is more difficult. As the evaluations of the defense (see~\autoref{figure:defense}) in~\autoref{section:defense} show, attacks against Item perform strongly even with the defense mechanism.
This is because Item has the simplest structure among the three recommendation algorithms.
In short, a recommender system established by a simple algorithm structure is usually more vulnerable to membership inference attacks.

\mypara{Distributions of Generated User Features.}
The combination of the dataset and recommender algorithm also matters. 
Higher attack performances can be obtained, when the distribution of user feature vectors generated by the shadow recommender system is more similar to the distribution generated by the target recommender system.
Trained with samples from a similar distribution, attack models are able to conduct an accurate inference.
Specifically, in~\autoref{figure:heatmap}, the attack of ``MLAI'' achieves a better performance than the one of ``ALLI'' (0.608 v.s. 0.547).
And we see from the visualization results in~\autoref{figure:specialcase_assumption3} that the above advantage comes from the smaller difference of feature distributions between the shadow recommender ``ML'' and the target recommender ``AI'' than the one between ``AL'' and ``LI''.
In summary, training data with distributions similar to the target data can boost the attack performances.

\section{Conclusion}

\label{section:conclusion}

Recommender systems have achieved tremendous success in real-world applications.
However, data used by recommender systems is highly sensitive.
In that case, successfully inferring a user's membership status from a target recommender may lead to severe privacy consequences.

In this paper, to investigate the privacy problem in recommender systems, we design various attack strategies of membership inference.
To the best of our knowledge, ours is the first work on the membership inference attacks against recommender systems.
Comparing to membership inference attacks on data sample-level classifiers, for recommender systems, our work focuses on the user-level membership status, which cannot be directly obtained from the system outputs.
To address these challenges, we propose a novel membership inference attack scheme, the core of which is to obtain user-level feature vectors based on the interactions between users and the target recommender, and input these feature vectors into attack models.
Extensive experiment results show the effectiveness and generalization ability of our attack.
To remedy the situation, we further propose a defense mechanism, namely Popularity Randomization.
Our empirical evaluations demonstrate that Popularity Randomization can largely mitigate the privacy risks.

\section*{Acknowledgements}
This work was supported by the Natural Science Foundation of China (61902219, 61972234, 62072279, 62102234), the Helmholtz Association within the project ``Trustworthy Federated Data Analytics'' (TFDA) (funding number ZT-I-OO1 4), the National Key R\&D Program of China with grant No. 2020YFB1406704, the Key Scientific and Technological Innovation Program of Shandong Province (2019JZZY010129), Shandong University multidisciplinary research and innovation team of young scholars (No. 2020QNQT017), and the Tencent WeChat Rhino-Bird Focused Research Program (JR-WXG2021411).
All content represents the opinion of the authors, which is not necessarily shared or endorsed by their respective employers and/or sponsors.


\bibliographystyle{plain}
\bibliography{normal_generated_py3}

\begin{thebibliography}{10}

\bibitem{BHPZ17}
Michael Backes, Mathias Humbert, Jun Pang, and Yang Zhang.
\newblock {walk2friends: Inferring Social Links from Mobility Profiles}.
\newblock In {\em {ACM SIGSAC Conference on Computer and Communications
  Security (CCS)}}, pages 1943--1957. ACM, 2017.

\bibitem{BWZZ17}
Ting Bai, Ji{-}Rong Wen, Jun Zhang, and Wayne~Xin Zhao.
\newblock {A Neural Collaborative Filtering Model with Interaction-based
  Neighborhood}.
\newblock In {\em {ACM International Conference on Information and Knowledge
  Management (CIKM)}}, pages 1979--1982. ACM, 2017.

\bibitem{BCMNSLGR13}
Battista Biggio, Igino Corona, Davide Maiorca, Blaine Nelson, Nedim Srndic,
  Pavel Laskov, Giorgio Giacinto, and Fabio Roli.
\newblock {Evasion Attacks against Machine Learning at Test Time}.
\newblock In {\em {European Conference on Machine Learning and Principles and
  Practice of Knowledge Discovery in Databases (ECML/PKDD)}}, pages 387--402.
  Springer, 2013.

\bibitem{CBK11}
Iv{\'a}n Cantador, Peter Brusilovsky, and Tsvi Kuflik.
\newblock {Second Workshop on Information Heterogeneity and Fusion in
  Recommender Systems (HetRec2011)}.
\newblock In {\em {ACM Conference on Recommender Systems (RecSys)}}, pages
  387--388. ACM, 2011.

\bibitem{CLEKS19}
Nicholas Carlini, Chang Liu, {\'U}lfar Erlingsson, Jernej Kos, and Dawn Song.
\newblock {The Secret Sharer: Evaluating and Testing Unintended Memorization in
  Neural Networks}.
\newblock In {\em {USENIX Security Symposium (USENIX Security)}}, pages
  267--284. USENIX, 2019.

\bibitem{CTWJHLRBSEOR20}
Nicholas Carlini, Florian Tram{\`{e}}r, Eric Wallace, Matthew Jagielski, Ariel
  Herbert{-}Voss, Katherine Lee, Adam Roberts, Tom~B. Brown, Dawn Song,
  {\'{U}}lfar Erlingsson, Alina Oprea, and Colin Raffel.
\newblock {Extracting Training Data from Large Language Models}.
\newblock {\em {CoRR abs/2012.07805}}, 2020.

\bibitem{CW17}
Nicholas Carlini and David Wagner.
\newblock {Towards Evaluating the Robustness of Neural Networks}.
\newblock In {\em {IEEE Symposium on Security and Privacy (S\&P)}}, pages
  39--57. IEEE, 2017.

\bibitem{CYZF20}
Dingfan Chen, Ning Yu, Yang Zhang, and Mario Fritz.
\newblock {GAN-Leaks: A Taxonomy of Membership Inference Attacks against
  Generative Models}.
\newblock In {\em {ACM SIGSAC Conference on Computer and Communications
  Security (CCS)}}, pages 343--362. ACM, 2020.

\bibitem{CCCR19}
Wanyu Chen, Fei Cai, Honghui Chen, and Maarten de~Rijke.
\newblock {Joint Neural Collaborative Filtering for Recommender Systems}.
\newblock {\em {ACM Transactions on Information Systems}}, 2019.

\bibitem{CTCP20}
Christopher A.~Choquette Choo, Florian Tram{\`e}r, Nicholas Carlini, and
  Nicolas Papernot.
\newblock {Label-Only Membership Inference Attacks}.
\newblock {\em {CoRR abs/2007.14321}}, 2020.

\bibitem{DK04}
Mukund Deshpande and George Karypis.
\newblock {Item-Based Top-N Recommendation Algorithms}.
\newblock {\em {ACM Transactions on Information Systems}}, 2004.

\bibitem{GDG17}
Tianyu Gu, Brendan Dolan-Gavitt, and Siddharth Grag.
\newblock {Badnets: Identifying Vulnerabilities in the Machine Learning Model
  Supply Chain}.
\newblock {\em {CoRR abs/1708.06733}}, 2017.

\bibitem{HK15}
F.~Maxwell Harper and Joseph~A Konstan.
\newblock {The MovieLens Datasets: History and Context}.
\newblock {\em {ACM Transactions on Interactive Intelligent Systems}}, 2015.

\bibitem{HMDC19}
Jamie Hayes, Luca Melis, George Danezis, and Emiliano~De Cristofaro.
\newblock {LOGAN: Evaluating Privacy Leakage of Generative Models Using
  Generative Adversarial Networks}.
\newblock {\em {Symposium on Privacy Enhancing Technologies Symposium}}, 2019.

\bibitem{HM16}
Ruining He and Julian McAuley.
\newblock {Ups and Downs: Modeling the Visual Evolution of Fashion Trends with
  One-Class Collaborative Filtering}.
\newblock In {\em {The Web Conference (WWW)}}, pages 507--517. ACM, 2016.

\bibitem{HLZNHC17}
Xiangnan He, Lizi Liao, Hanwang Zhang, Liqiang Nie, Xia Hu, and Tat{-}Seng
  Chua.
\newblock {Neural Collaborative Filtering}.
\newblock In {\em {International Conference on World Wide Web (WWW)}}, pages
  173--182. ACM, 2017.

\bibitem{HZKC16}
Xiangnan He, Hanwang Zhang, Min{-}Yen Kan, and Tat{-}Seng Chua.
\newblock {Fast Matrix Factorization for Online Recommendation with Implicit
  Feedback}.
\newblock In {\em {International ACM SIGIR Conference on Research and
  Development in Information Retrieval (SIGIR)}}, pages 549--558. ACM, 2016.

\bibitem{HJBGZ21}
Xinlei He, Jinyuan Jia, Michael Backes, Neil~Zhenqiang Gong, and Yang Zhang.
\newblock {Stealing Links from Graph Neural Networks}.
\newblock In {\em {USENIX Security Symposium (USENIX Security)}}. USENIX, 2021.

\bibitem{HKR00}
Jonathan~L. Herlocker, Joseph~A. Konstan, and John Riedl.
\newblock {Explaining Collaborative Filtering Recommendations}.
\newblock In {\em {ACM Conference on Computer Supported Cooperative Work
  (CSCW)}}, pages 241--250. ACM, 2000.

\bibitem{JCBKP20}
Matthew Jagielski, Nicholas Carlini, David Berthelot, Alex Kurakin, and Nicolas
  Papernot.
\newblock {High Accuracy and High Fidelity Extraction of Neural Networks}.
\newblock In {\em {USENIX Security Symposium (USENIX Security)}}, pages
  1345--1362. USENIX, 2020.

\bibitem{JOBLNL18}
Matthew Jagielski, Alina Oprea, Battista Biggio, Chang Liu, Cristina
  Nita-Rotaru, and Bo~Li.
\newblock {Manipulating Machine Learning: Poisoning Attacks and Countermeasures
  for Regression Learning}.
\newblock In {\em {IEEE Symposium on Security and Privacy (S\&P)}}, pages
  19--35. IEEE, 2018.

\bibitem{JSBZG19}
Jinyuan Jia, Ahmed Salem, Michael Backes, Yang Zhang, and Neil~Zhenqiang Gong.
\newblock {MemGuard: Defending against Black-Box Membership Inference Attacks
  via Adversarial Examples}.
\newblock In {\em {ACM SIGSAC Conference on Computer and Communications
  Security (CCS)}}, pages 259--274. ACM, 2019.

\bibitem{K01}
George Karypis.
\newblock {Evaluation of Item-Based Top-N Recommendation Algorithms}.
\newblock In {\em {ACM International Conference on Information and Knowledge
  Management (CIKM)}}, pages 247--254. ACM, 2001.

\bibitem{K08}
Yehuda Koren.
\newblock {Factorization Meets the Neighborhood: a Multifaceted Collaborative
  Filtering Model}.
\newblock In {\em {ACM Conference on Knowledge Discovery and Data Mining
  (KDD)}}, pages 426--434. ACM, 2008.

\bibitem{K09}
Yehuda Koren.
\newblock {Collaborative Filtering with Temporal Dynamics}.
\newblock In {\em {ACM Conference on Knowledge Discovery and Data Mining
  (KDD)}}, pages 447--456. ACM, 2009.

\bibitem{LF20}
Klas Leino and Matt Fredrikson.
\newblock {Stolen Memories: Leveraging Model Memorization for Calibrated
  White-Box Membership Inference}.
\newblock In {\em {USENIX Security Symposium (USENIX Security)}}, pages
  1605--1622. USENIX, 2020.

\bibitem{LHZG19}
Zheng Li, Chengyu Hu, Yang Zhang, and Shanqing Guo.
\newblock {How to Prove Your Model Belongs to You: A Blind-Watermark based
  Framework to Protect Intellectual Property of DNN}.
\newblock In {\em {Annual Computer Security Applications Conference (ACSAC)}},
  pages 126--137. ACM, 2019.

\bibitem{LZ21}
Zheng Li and Yang Zhang.
\newblock {Membership Leakage in Label-Only Exposures}.
\newblock In {\em {ACM SIGSAC Conference on Computer and Communications
  Security (CCS)}}. ACM, 2021.

\bibitem{MSCS19}
Luca Melis, Congzheng Song, Emiliano~De Cristofaro, and Vitaly Shmatikov.
\newblock {Exploiting Unintended Feature Leakage in Collaborative Learning}.
\newblock In {\em {IEEE Symposium on Security and Privacy (S\&P)}}, pages
  497--512. IEEE, 2019.

\bibitem{NSH18}
Milad Nasr, Reza Shokri, and Amir Houmansadr.
\newblock {Machine Learning with Membership Privacy using Adversarial
  Regularization}.
\newblock In {\em {ACM SIGSAC Conference on Computer and Communications
  Security (CCS)}}, pages 634--646. ACM, 2018.

\bibitem{NSH19}
Milad Nasr, Reza Shokri, and Amir Houmansadr.
\newblock {Comprehensive Privacy Analysis of Deep Learning: Passive and Active
  White-box Inference Attacks against Centralized and Federated Learning}.
\newblock In {\em {IEEE Symposium on Security and Privacy (S\&P)}}, pages
  1021--1035. IEEE, 2019.

\bibitem{NSTPC21}
Milad Nasr, Shuang Song, Abhradeep Thakurta, Nicolas Papernot, and Nicholas
  Carlini.
\newblock {Adversary Instantiation: Lower Bounds for Differentially Private
  Machine Learning}.
\newblock In {\em {IEEE Symposium on Security and Privacy (S\&P)}}. IEEE, 2021.

\bibitem{PMSW18}
Nicolas Papernot, Patrick McDaniel, Arunesh Sinha, and Michael Wellman.
\newblock {SoK: Towards the Science of Security and Privacy in Machine
  Learning}.
\newblock In {\em {IEEE European Symposium on Security and Privacy (Euro
  S\&P)}}, pages 399--414. IEEE, 2018.

\bibitem{PSMRTE18}
Nicolas Papernot, Shuang Song, Ilya Mironov, Ananth Raghunathan, Kunal Talwar,
  and {\'{U}}lfar Erlingsson.
\newblock {Scalable Private Learning with {PATE}}.
\newblock In {\em {International Conference on Learning Representations
  (ICLR)}}, 2018.

\bibitem{PB07}
Michael~J. Pazzani and Daniel Billsus.
\newblock {Content-Based Recommendation Systems}.
\newblock In {\em {The Adaptive Web, Methods and Strategies of Web
  Personalization}}, pages 325--341. Springer, 2007.

\bibitem{PD05}
Huseyin Polat and Wenliang Du.
\newblock {SVD-based Collaborative Filtering with Privacy}.
\newblock In {\em {ACM Symposium on Applied Computing (SAC)}}, pages 791--795.
  ACM, 2005.

\bibitem{SDSOJ19}
Alexandre Sablayrolles, Matthijs Douze, Cordelia Schmid, Yann Ollivier, and
  Herv{\'e} J{\'e}gou.
\newblock {White-box vs Black-box: Bayes Optimal Strategies for Membership
  Inference}.
\newblock In {\em {International Conference on Machine Learning (ICML)}}, pages
  5558--5567. PMLR, 2019.

\bibitem{SM07}
Ruslan Salakhutdinov and Andriy Mnih.
\newblock {Probabilistic Matrix Factorization}.
\newblock In {\em {Annual Conference on Neural Information Processing Systems
  (NIPS)}}, pages 1257--1264. NIPS, 2007.

\bibitem{SZHBFB19}
Ahmed Salem, Yang Zhang, Mathias Humbert, Pascal Berrang, Mario Fritz, and
  Michael Backes.
\newblock {ML-Leaks: Model and Data Independent Membership Inference Attacks
  and Defenses on Machine Learning Models}.
\newblock In {\em {Network and Distributed System Security Symposium (NDSS)}}.
  Internet Society, 2019.

\bibitem{SKKR01}
Badrul~Munir Sarwar, George Karypis, Joseph~A. Konstan, and John Riedl.
\newblock {Item-Based Collaborative Filtering Recommendation Algorithms}.
\newblock In {\em {International Conference on World Wide Web (WWW)}}, pages
  285--295. ACM, 2001.

\bibitem{SFHS07}
J.~Ben Schafer, Dan Frankowski, Jon Herlocker, and Shilad Sen.
\newblock {Collaborative Filtering Recommender Systems}.
\newblock In {\em {The Adaptive Web, Methods and Strategies of Web
  Personalization}}, pages 291--324. Springer, 2007.

\bibitem{SH21}
Virat Shejwalkar and Amir Houmansadr.
\newblock {Membership Privacy for Machine Learning Models Through Knowledge
  Transfer}.
\newblock In {\em {AAAI Conference on Artificial Intelligence (AAAI)}}. AAAI,
  2021.

\bibitem{SSSS17}
Reza Shokri, Marco Stronati, Congzheng Song, and Vitaly Shmatikov.
\newblock {Membership Inference Attacks Against Machine Learning Models}.
\newblock In {\em {IEEE Symposium on Security and Privacy (S\&P)}}, pages
  3--18. IEEE, 2017.

\bibitem{STBH11}
Reza Shokri, Georgios Theodorakopoulos, Jean-Yves~Le Boudec, and Jean-Pierre
  Hubaux.
\newblock {Quantifying Location Privacy}.
\newblock In {\em {IEEE Symposium on Security and Privacy (S\&P)}}, pages
  247--262. IEEE, 2011.

\bibitem{SS19}
Congzheng Song and Vitaly Shmatikov.
\newblock {Auditing Data Provenance in Text-Generation Models}.
\newblock In {\em {ACM Conference on Knowledge Discovery and Data Mining
  (KDD)}}, pages 196--206. ACM, 2019.

\bibitem{SWW18}
Peijie Sun, Le~Wu, and Meng Wang.
\newblock {Attentive Recurrent Social Recommendation}.
\newblock In {\em {International ACM SIGIR Conference on Research and
  Development in Information Retrieval (SIGIR)}}, pages 185--194. ACM, 2018.

\bibitem{SWZFHW20}
Peijie Sun, Le~Wu, Kun Zhang, Yanjie Fu, Richang Hong, and Meng Wang.
\newblock {Dual Learning for Explainable Recommendation: Towards Unifying User
  Preference Prediction and Review Generation}.
\newblock In {\em {The Web Conference (WWW)}}, pages 837--847. ACM, 2020.

\bibitem{TKPGBM17}
Florian Tram{\`e}r, Alexey Kurakin, Nicolas Papernot, Ian Goodfellow, Dan
  Boneh, and Patrick McDaniel.
\newblock {Ensemble Adversarial Training: Attacks and Defenses}.
\newblock In {\em {International Conference on Learning Representations
  (ICLR)}}, 2017.

\bibitem{TZJRR16}
Florian Tram{\`e}r, Fan Zhang, Ari Juels, Michael~K. Reiter, and Thomas
  Ristenpart.
\newblock {Stealing Machine Learning Models via Prediction APIs}.
\newblock In {\em {USENIX Security Symposium (USENIX Security)}}, pages
  601--618. USENIX, 2016.

\bibitem{MH08}
Laurens van~der Maaten and Geoffrey Hinton.
\newblock {Visualizing Data using t-SNE}.
\newblock {\em {Journal of Machine Learning Research}}, 2008.

\bibitem{WF20}
Bogdan Walek and Vladimir Fojtik.
\newblock {A Hybrid Recommender System for Recommending Relevant Movies Using
  An Expert System}.
\newblock {\em {Expert Systems with Applications}}, 2020.

\bibitem{YGFJ18}
Samuel Yeom, Irene Giacomelli, Matt Fredrikson, and Somesh Jha.
\newblock {Privacy Risk in Machine Learning: Analyzing the Connection to
  Overfitting}.
\newblock In {\em {IEEE Computer Security Foundations Symposium (CSF)}}, pages
  268--282. IEEE, 2018.

\end{thebibliography}

\appendix

\section{Defense}
In this section, we demonstrate the defense methodology in detail (Appendix \ref{section:defensemethodology}), discuss about visualization results (Appendix \ref{section:defenseresults}), and analyze the impacts of Popularity Randomization on original recommendation performances (Appendix \ref{section: Impacts on Recommendation Performances}). 

\begin{figure*}[!t]
\centering
\begin{subfigure}{\columnwidth}
\centering
\includegraphics[width=0.45\columnwidth]{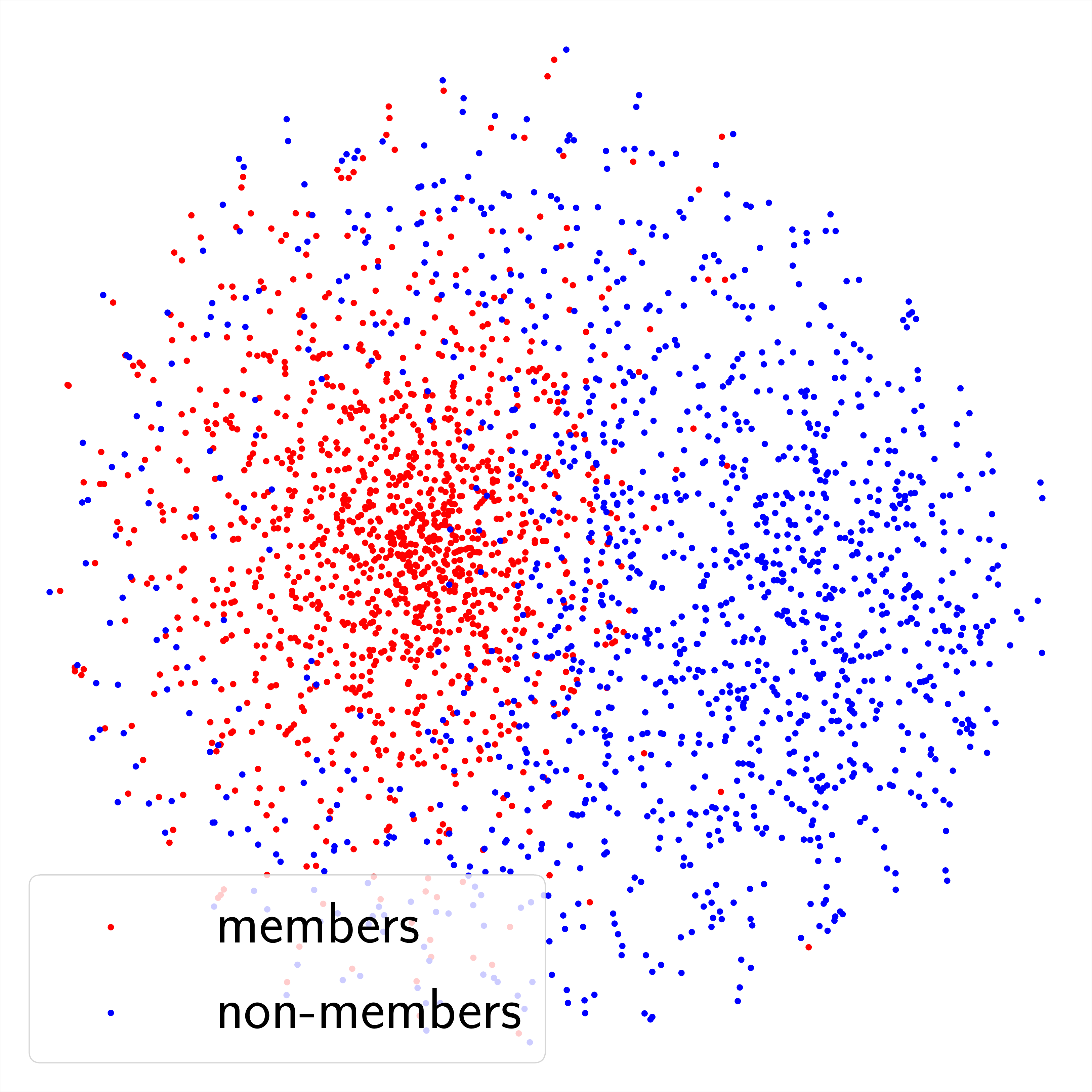}
\includegraphics[width=0.45\columnwidth]{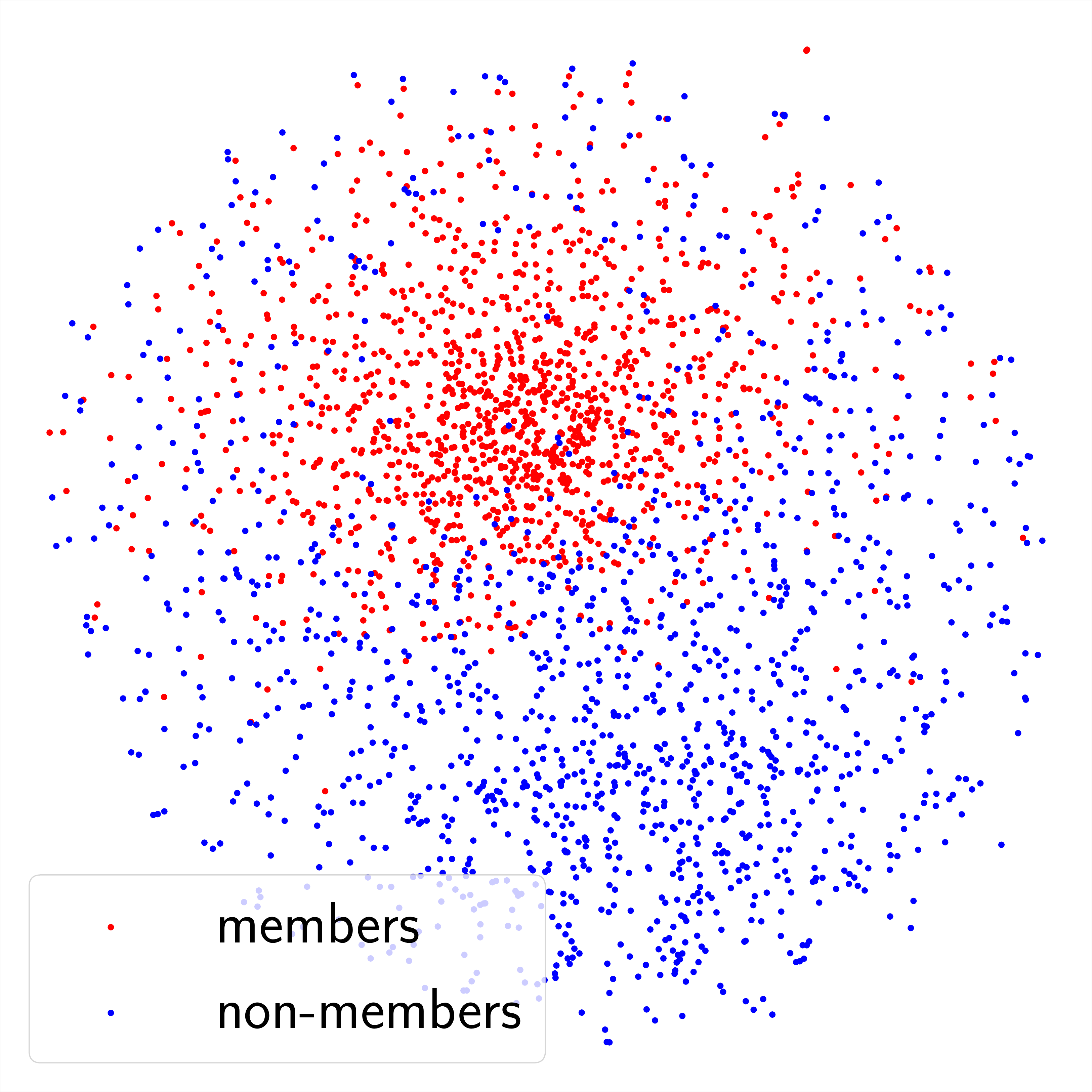}
\caption{ADM\_Item\_shadow (left) and ADM\_Item\_target (right) with the defense mechanism.}
\label{figure:tsne_amazon_itembase_random}
\end{subfigure}
\begin{subfigure}{\columnwidth}
\centering
\includegraphics[width=0.45\columnwidth]{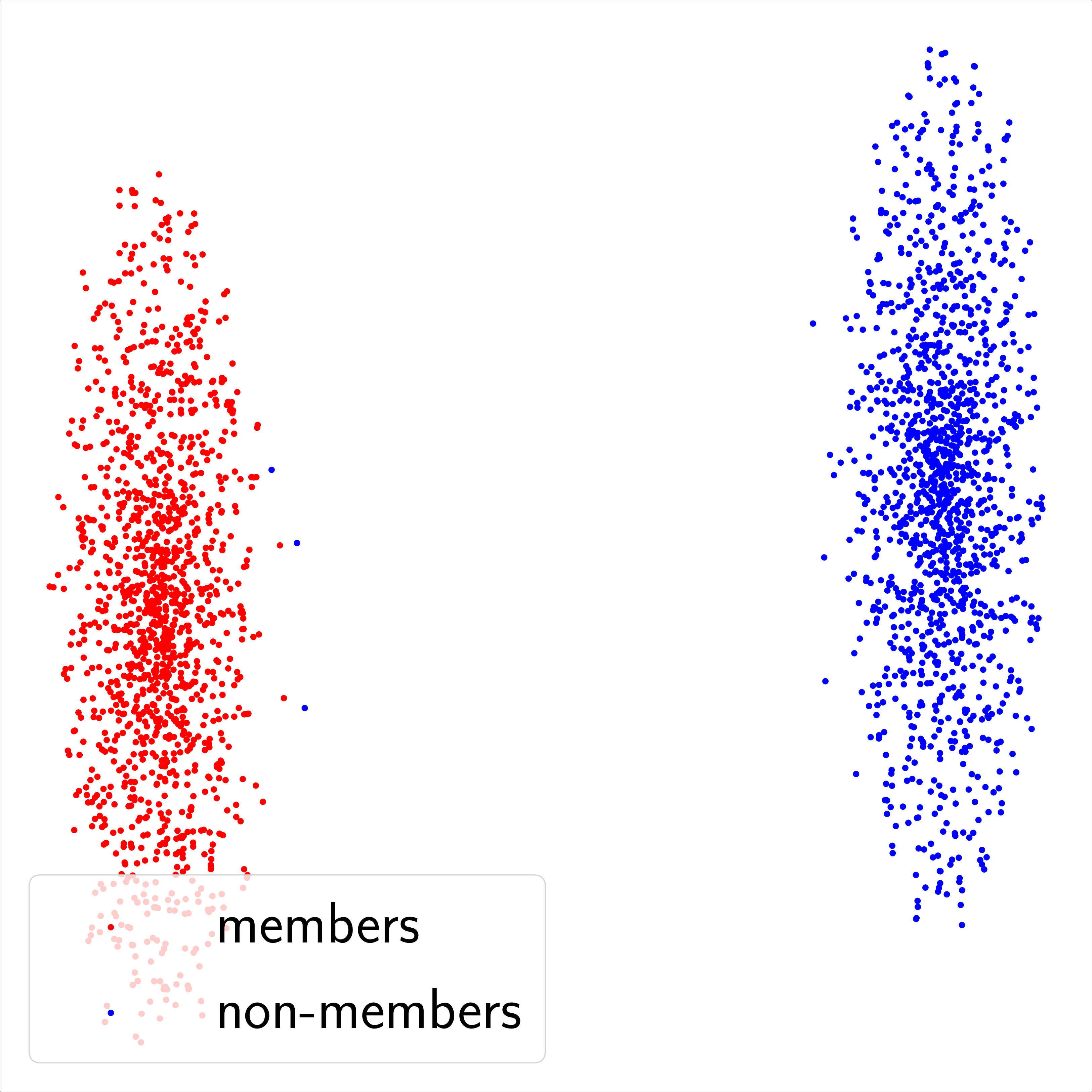}
\includegraphics[width=0.45\columnwidth]{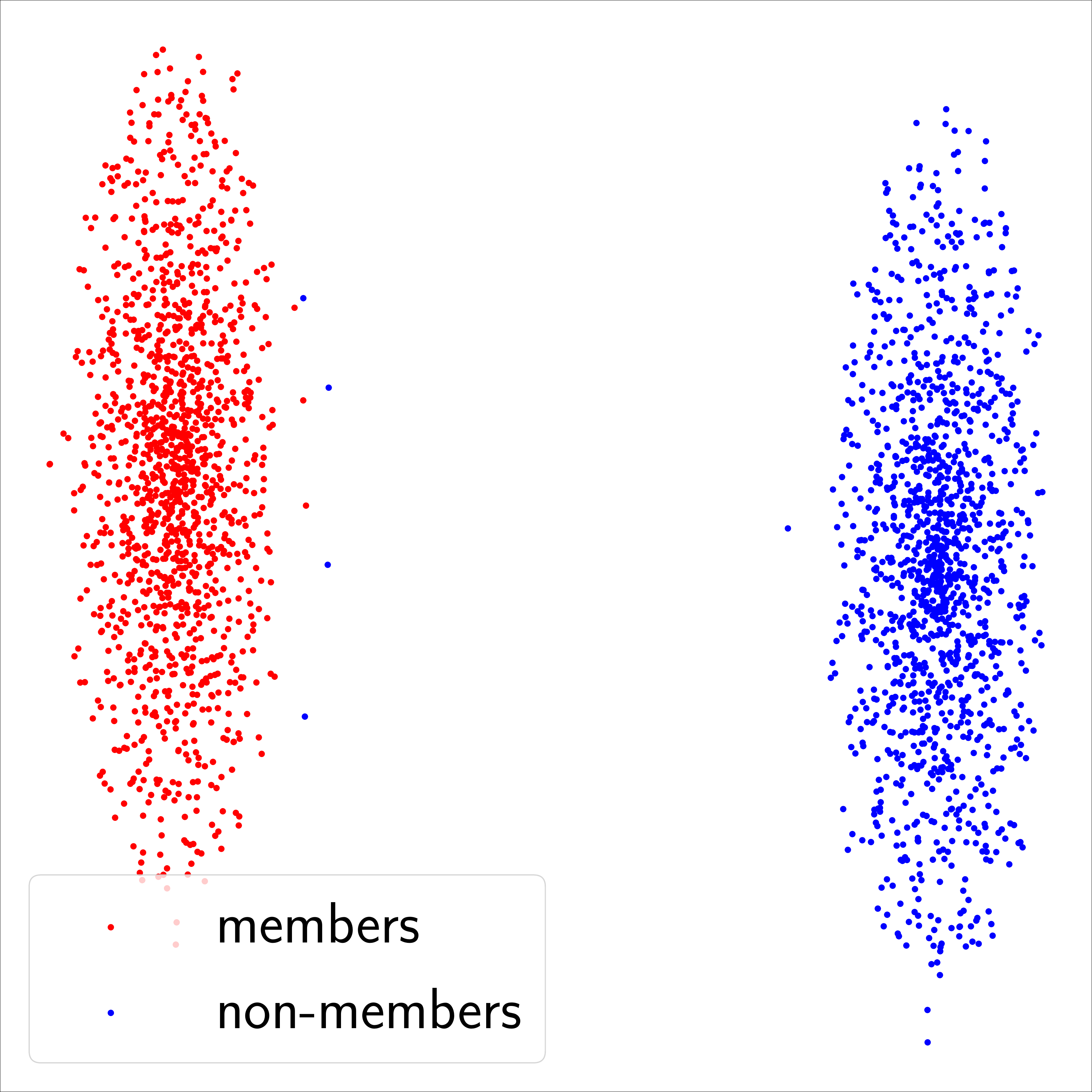}
\caption{ADM\_Item\_shadow (left) and ADM\_Item\_target (right) without the defense mechanism.}
\label{figure:tsne_amazon_itembase}
\end{subfigure}
\begin{subfigure}{\columnwidth}
\centering
\includegraphics[width=0.45\columnwidth]{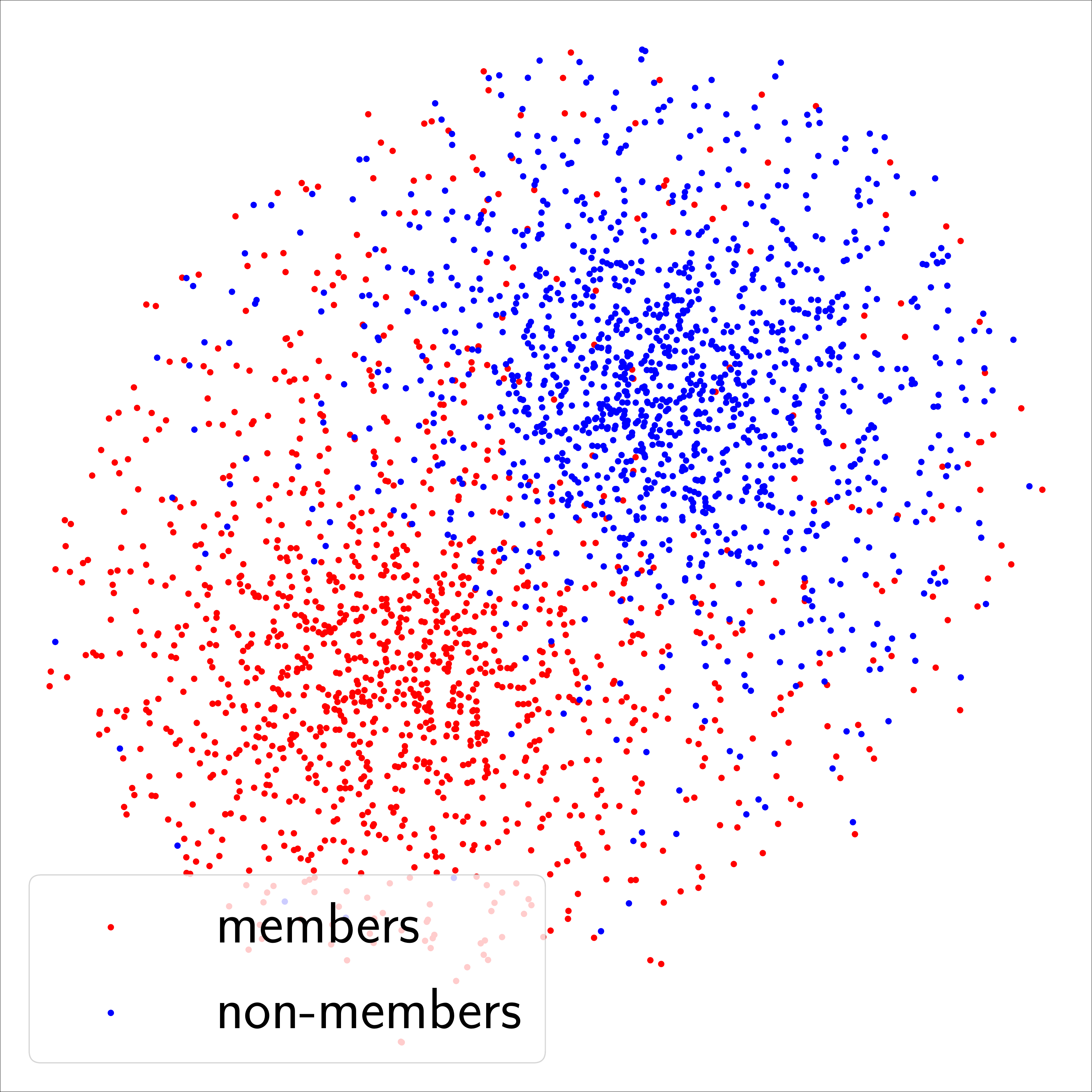}
\includegraphics[width=0.45\columnwidth]{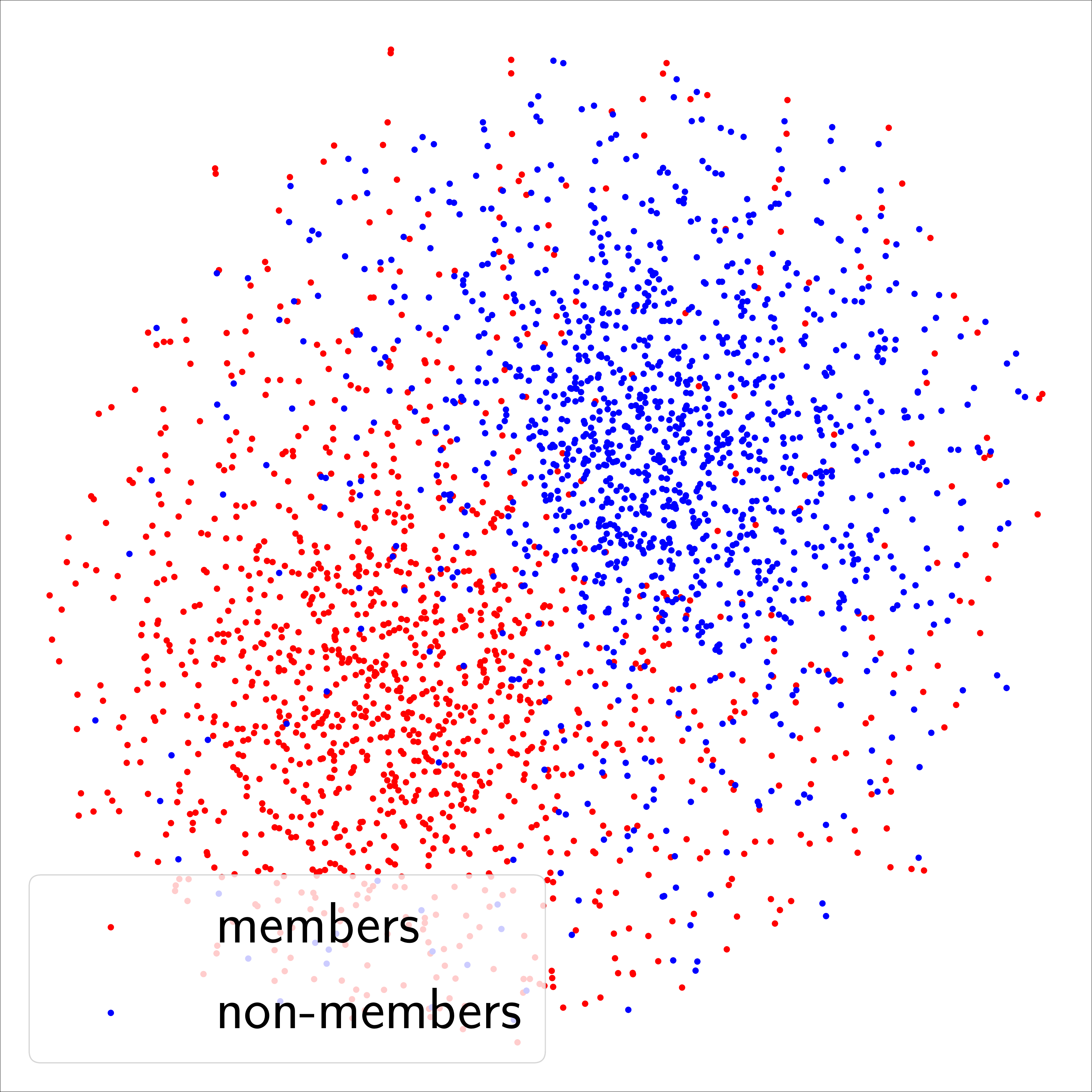}
\caption{ADM\_LFM\_shadow (left) and ADM\_LFM\_target (right) with the defense mechanism.}
\label{figure:tsne_amazon_lfm_random}
\end{subfigure}
\begin{subfigure}{\columnwidth}
\centering
\includegraphics[width=0.45\columnwidth]{pic/ADM_LFM_shadow.pdf}
\includegraphics[width=0.45\columnwidth]{pic/ADM_LFM_target.pdf}
\caption{ADM\_LFM\_shadow (left) and ADM\_LFM\_target (right) without the defense mechanism.}
\label{figure:tsne_amazon_lfm}
\end{subfigure}
\caption{Data distributions by t-SNE, in which red points represent members and blue points denote non-member.
(a) Data points in the shadow and target recommenders using Item on the ADM dataset with the defense mechanism.
(b) Data points in the shadow and target recommenders using Item on the ADM dataset without the defense mechanism.
(c) Data points in the shadow and target recommenders using LFM on the ADM dataset with the defense mechanism.
(d) Data points in the shadow and target recommenders using LFM on the ADM dataset without the defense mechanism.}
\label{figure:tsne_defense}
\end{figure*} 

\subsection{Methodology}
\label{section:defensemethodology}

In the original setting in~\autoref{section:MIA}, non-members are provided with the most popular items.
However, the most popular items for different users are the same, resulting in that feature vectors of non-members are extremely similar and easily distinguished.
To address this problem, an intuitive way is to increase the randomness of non-members' recommendations. In the paper, we propose a defense mechanism named \emph{Popularity Randomization}. Specifically, a larger number of the most popular items are first selected as candidates.
Then, we conduct a random selection, i.e., we randomly pick 10\% of candidates as recommendations for non-members.
Note that all recommendations for non-members are still the most popular items.
In that case, recommendations to different non-members are diverse and thus the similarities among non-members' feature vectors are decreased.
The above defense strategy can be formulated as follows:

\begin{figure}[!ht]
\centering
\includegraphics[width=0.95\columnwidth]{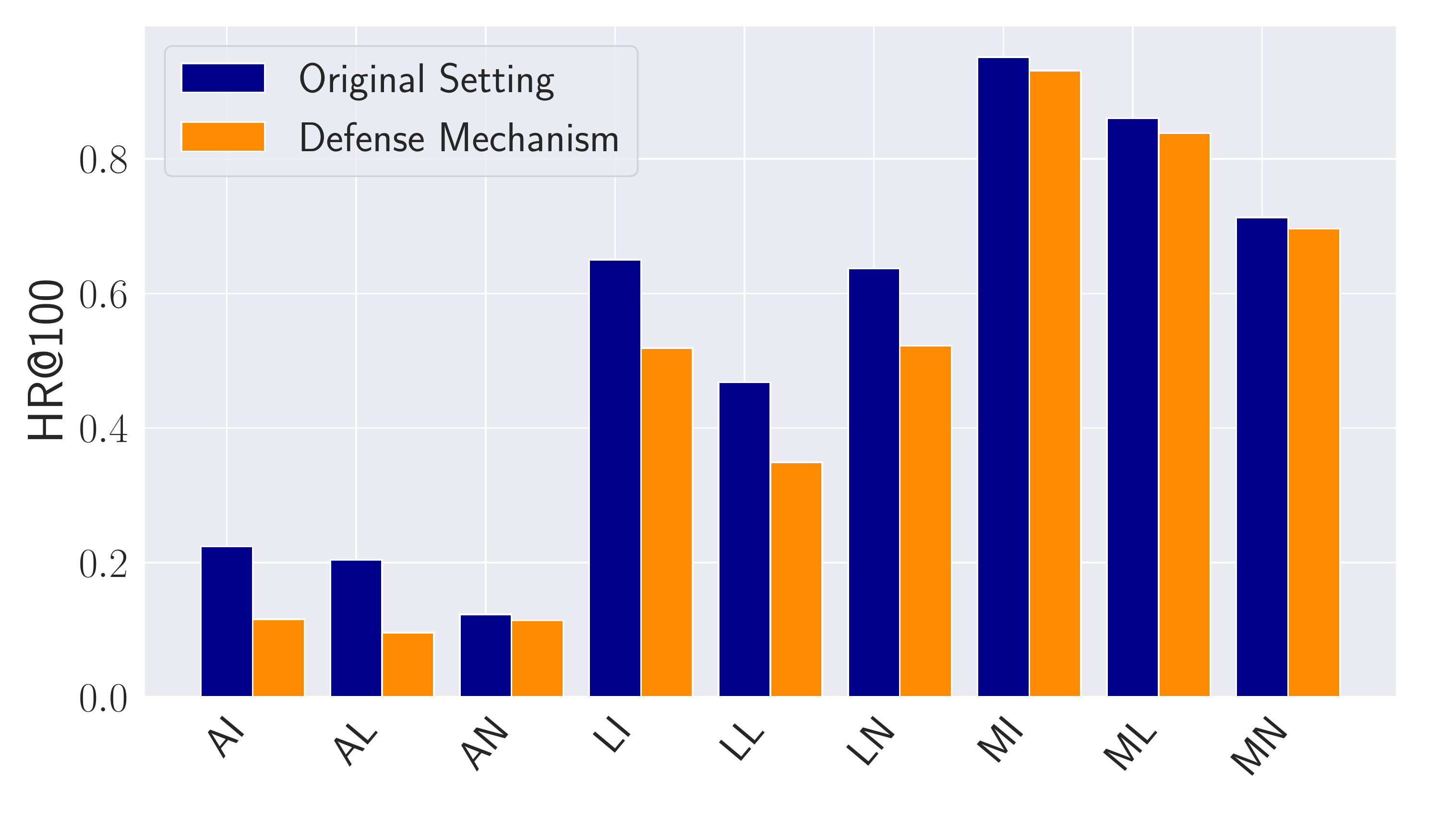}
\caption{Comparisons of recommendation performances before against after deploying the defense mechanism.}
\label{figure:eval_defense}
\end{figure}

First, a recommender system $\mathcal{A}_{RS}$ sorts all items by popularity, i.e., 
$$\mathcal{A}_{RS}: \mathcal{D}_{original} \stackrel{Sort}{\longrightarrow} \mathcal{S}_{sorted},$$
where $\mathcal{D}_{original}$ is the original dataset and $\stackrel{Sort}{\longrightarrow}$ is to sort items by popularity.
And $\mathcal{S}_{sorted}$ is an ordered sequence, including all items from $\mathcal{D}_{original}$.
In $\mathcal{S}_{sorted}$, higher items are more popular.

Second, based on the number of recommendations and the pre-set ratio, the recommender system selects a number of the most popular items as candidates, which can be defined as the following functions:
$$\alpha_{PR} = \frac{N_{rec}}{N_{cand}},$$
$$\mathcal{S}_{cand} = \{x | x \in \mathcal{S}_{sorted}, Ind_{x} \leq N_{cand}\},$$
where $\alpha_{PR}$ is the ratio of recommendations to candidates.
$\mathcal{S}_{cand}$ is a sequence containing all candidates selected by the recommender system from $\mathcal{S}_{sorted}$.
$N_{rec}$ and $N_{cand}$ are the numbers of the recommendations and the candidates.
And $Ind_{x}$ is the index of $x$ in $\mathcal{S}_{sorted}$.

Finally, non-members are randomly provided with items from candidates as recommendations, which can be defined as the following function:
$$f(\mathcal{S}_{cand}) = \mathcal{R}^{out} \quad (|\mathcal{R}^{out}| = N_{rec}),$$
where $f(\mathcal{S}_{cand})$ is to randomly select recommendations from $\mathcal{S}_{cand}$.
And $|\mathcal{R}^{out}|$, which is equal to $N_{rec}$, is the number of recommendations to non-members.

\subsection{Visualization Results}
\label{section:defenseresults}

In this section, we visualize user feature vectors by t-SNE to show the differences of the distributions between members and non-members in the shadow and target datasets.
In~\autoref{figure:tsne_defense}, the red points represent feature vectors of members and the blue points denote feature vectors of non-members.
We conclude from the comparison between~\autoref{figure:tsne_amazon_itembase_random} and~\autoref{figure:tsne_amazon_itembase} that Popularity Randomization can decrease the differences between the feature vectors of members and non-members, and hinder the attack performance considerably.
We find a similar phenomenon in~\autoref{figure:tsne_amazon_lfm_random} and~\autoref{figure:tsne_amazon_lfm}.
These visualization results show the effectiveness of Popularity Randomization.

As~\autoref{figure:defense} shows, with the defense mechanism, the attack against LFM can only reach comparable performances with Random Guess.
In contrast, when the target algorithm is Item, the attack achieves strong performances. To explain this, as~\autoref{figure:tsne_amazon_itembase_random} and~\autoref{figure:tsne_amazon_lfm_random} show, the areas of the red and blue points rarely overlap when using Item.
However, there exist many red points in the area of the blue points when using LFM, leading to a poor attack performance.

\subsection{Impacts on Recommendation Performances}
\label{section: Impacts on Recommendation Performances}

We next investigate the impacts on the performance of the target recommender with Popularity Randomization.
In~\autoref{figure:eval_defense}, the results show that the performances of the recommender only slightly drops.
For instance, when the target recommender uses Item on the ml-1m dataset, Popularity Randomization only achieves a $2\%$ drop in the recommendation performance. Therefore, our proposed defense strategy can mitigate the attack risk effectively while preserving the original recommendation performances.


\end{document}